\documentclass[manuscript]{acmart}
\usepackage{amsmath}  % if needed for math environments
\usepackage{array}
\usepackage{booktabs}
\usepackage{courier}
\usepackage{enumitem}
\usepackage{fontawesome}
\usepackage{graphicx} % for \includegraphics
\usepackage{helvet}
\usepackage{hyperref}
\usepackage[utf8]{inputenc}
\usepackage{listings}
\usepackage{microtype}
\usepackage{multirow}
\usepackage{placeins}  % Ensures floats do not move past a section
\usepackage{subcaption}
\usepackage{tcolorbox}
\usepackage{times}
\usepackage{url}
\usepackage{xargs}
\usepackage{xcolor}
\usepackage{xspace}

\newcommand{\td}[1] % To do macro
{{\color{red} $\leftarrow$ {\bf TO DO}: #1 }}
\newcommand{\ms}[1]{{\footnotesize\texttt{#1}}} % monospace macro
\definecolor{red}{RGB}{178,34,34}
\definecolor{lightgray}{HTML}{F7F7F7}

\tcbset{
  llmoutput/.style={
    enhanced,
    boxrule=0.6pt,
    colback=lightgray,
    colframe=black!35,
    arc=2mm,
    left=2.2mm,right=2.2mm,top=1.6mm,bottom=1.6mm,
  }
}

\lstdefinestyle{prompt}{
  basicstyle=\ttfamily\small,
  breaklines=true,
  frame=single,
  backgroundcolor=\color{gray!10},
}

\AtBeginDocument{%
  }

\setcopyright{acmlicensed}
\copyrightyear{2018}
\acmYear{2018}
\acmDOI{XXXXXXX.XXXXXXX}
\acmConference[Conference acronym 'XX]{Make sure to enter the correct
  conference title from your rights confirmation email}{June 03--05,
  2018}{Woodstock, NY}
\acmISBN{978-1-4503-XXXX-X/2018/06}

\begin{document}

\title{Deconstructing Taste: Toward a Human-Centered AI Framework for Modeling Consumer Aesthetic Perceptions}

%Matthew K. Hong, Joey Li, Alexandre Filipowicz, Monica Van, Kalani Murakami, Yan-Ying Chen, Shiwali Mohan, Shabnam Hakimi, Matthew Klenk
\author{Matthew K. Hong}
\author{Joey Li}
\author{Alexandre Filipowicz}
\author{Monica Van}
\author{Kalani Murakami}
\author{Yan-Ying Chen}
\author{Shiwali Mohan}
\author{Shabnam Hakimi}
\author{Matthew Klenk}
\affiliation{%
  \institution{Toyota Research Institute}
  \city{Los Altos}
  \state{California}
  \country{USA}
}
\email{{matt.hong, joey.li, alex.filipowicz, monica.van, kalani.murakami, yan-ying.chen, shiwali.mohan.ctr, shabnam.hakimi, matt.klenk}@tri.global}

\begin{abstract}
Understanding and modeling consumers' stylistic taste such as "sporty" is crucial for creating designs that truly connect with target audiences. However, capturing taste during the design process remains challenging because taste is abstract and subjective, and preference data alone provides limited guidance for concrete design decisions. This paper proposes an integrated human-centered computational framework that links subjective evaluations (e.g., perceived luxury of car wheels) with domain-specific features (e.g., spoke configuration) and computer vision-based measures (e.g., texture). By jointly modeling human-derived (consumer and designer) and machine-extracted features, our framework advances aesthetic assessment by explicitly linking model outcomes to interpretable design features. In particular, it demonstrates how perceptual features, domain-specific design patterns, and consumers' own interpretations of style contribute to aesthetic evaluations. This framework will enable product teams to better understand, communicate, and critique aesthetic decisions, supporting improved anticipation of consumer taste and more informed exploration of design alternatives at design time.
\end{abstract}






\maketitle

\section{Introduction}
%Computational creativity research aims to design algorithms, tools, and frameworks that augment human creativity~\cite{colton2012computational}.
Designing aesthetic forms that resonate with consumers is a central challenge in product design, as designers must navigate the tension between creative expression and the constraints of high production costs and lengthy engineering lead times. Computationally modeling consumer aesthetic taste offers a way to address this challenge by enabling rapid, informative feedback throughout the creative process, allowing designers to quickly iterate and prioritize actionable insights that are more likely to produce products that appeal to target audiences and drive business impact. 

Prior studies in the automotive industry have linked aesthetic appeal to 60\% of purchase decisions~\cite{kreuzbauer2005embodied}. Yet, improving aesthetic appeal for vehicle products requires costly upfront investment amounting to \$3 billion for major redesigns~\cite{burnap2023product}. Advanced AI techniques, including generative adversarial networks (GANs) and other deep learning models, are emerging as powerful tools for increasing the scope of AI's role in these high-stakes scenarios~\cite{burnap2023product}. However, the subjective, time-variant, and context-dependent nature of consumers' taste poses challenges in ensuring that algorithmic recommendations for design consistently align with diverse and evolving consumer expectations. This state reflects a broader need to design human-AI co-creative systems that model aesthetic perceptions of style in ways that can be meaningfully communicated back to designers---in the context of their own design language---to support an adaptive human-centered design process.
%This challenge is particularly salient in automotive design where engineering considerations cause long lead times between styling decisions and product launches. 

Traditionally, assessments of consumer taste relied on human-derived evaluations---for instance, expert critiques and consumer surveys in design clinics---to gauge style descriptors like \textit{sporty} or \textit{luxurious}. However, their subjective nature can make it difficult to translate consumer insights into precise design interventions, leading to costly redesigns to ensure consumers' expectations are met\footnote{A single design clinic for a new vehicle can cost over \$100,000, with annual expenses for a single manufacturer reaching tens of millions of dollars~\cite{burnap2023product}}. Recently, large scale preference datasets and their reward model offer scalable means to capture consumer taste by enabling preference-based model post-training pipelines~\cite{zhang2024learning,xu2023imagereward,kirstain2023pick,liang2025aesthetic}. While whole-image preference ratings indicate relative desirability, they however do not encode which visual attributes, semantic cues, or design patterns drive those judgments. For design practice, this means models trained on image-level preferences can reproduce stylistic tendencies without exposing the principles that designers need to critique, adapt, or intentionally subvert them.

%However, text prompts are often unintelligible to designers, and models trained on aesthetic preference ratings produce homogenized outputs that converge on polished, clean and curved forms perceived as overly AI-like.

Within the machine learning community, researchers have recognized the importance of combining high-level semantic attributes with multiple low-level hand-crafted features (e.g., rule of thirds, spatial distribution of edges) and generic image features (e.g., texture, saturation of colors) for shaping consumer perceptions of overall attractiveness~\cite{brachmann2017computational, veryzer1993aesthetic}. With the emergence of more sophisticated computer vision techniques and powerful vision language models, it is now possible to comprehensively model product aesthetics through a range of low-level visual features such as texture, spatial composition, and color distribution, and high-level semantic features that reflect both designers' and consumers' understanding of designed artifacts~\cite{machajdik2010affective, brachmann2017computational}. For instance, low-level features extracted automatically from images can provide objective measurements of visual form, while high-level, human-interpretable features offer fine-grained semantic distinctions. Together, these complementary representations help disentangle context-dependent relationships between these features, enabling more informed creative decision-making by revealing not only what consumers prefer, but also why those preferences arise across contexts~\cite{crilly2004seeing}. How then to effectively represent aesthetic features in a system that supports design creativity remains a critical open question.

Building on these developments, this paper proposes an integrated framework for aligning computational models of aesthetics with the nuanced ways in which humans\textemdash both designers and consumers\textemdash perceive and evaluate design elements. Our work examines how integrating consumers' subjective judgments (e.g., perceived luxury of car wheels), designer-informed visual features (e.g., spoke configuration), and machine-extracted visual features (e.g., texture or spatial composition) yields deeper understandings of consumer taste. In doing so, we address a significant gap in the AI and design literature: while many existing methods excel at generating or optimizing designs toward aesthetic goals, relatively few provide insights into why certain designs might resonate more effectively with target audiences~\cite{zhou2021evaluation,burnap2023product}. This research contributes to both theory and practice by 1) linking psychological models of aesthetic response with computational techniques and 2) proposing a generalizable, user-centered approach for early product development. 

To this end, we provide an example study that demonstrates how this framework can be used to assess people's perception of product aesthetics (specifically, vehicle wheel aesthetics). This study focuses on the two following research questions:
%\vspace{1em}

\noindent\\ \textbf{RQ1}: What features contribute most to people's perception of aesthetic style? \\
\textbf{RQ2}: Do people form differentiable perceptions of aesthetic style?
\vspace{1em}

%Through this integrated method, cross-functional design teams can make informed predictions about consumers' taste, refining both product ideation and development processes early and also toward later stages of design that more reliably achieve the desired emotional and perceptual impact. 

%By bridging this gap, our proposed framework aims to inform data-driven design decisions and augment designers' intuition with insights developed early in the product design phase. Specifically, we demonstrate how integrating subjective and objective features yields deeper understandings of consumer taste, ultimately enabling more strategic, AI-assisted design generation. 

%What makes this human-centered?
%Designer informed categories + consumer judgements + machine extracted features
\section{Background and Related Work}
We position our work within creativity literature in Psychology and Computation to inform a holistic understanding of product aesthetics. In doing so, we highlight existing gaps in the literature that motivate our research questions.

\begin{figure}[t]  % 't' for top placement
\centering
\includegraphics[width=\linewidth]{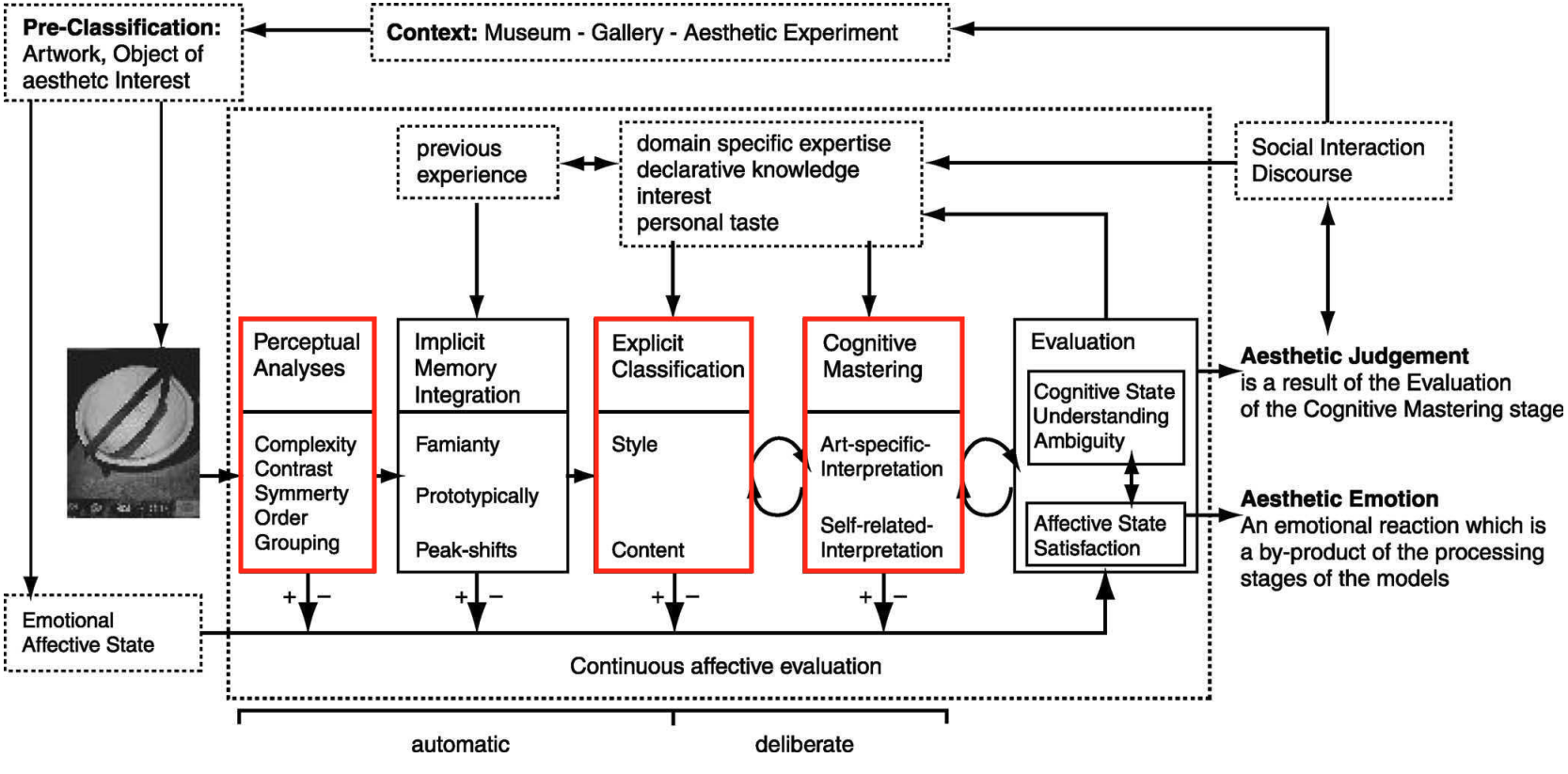}
\caption{Model of aesthetic experiences (taken from~\cite{leder2004model}). The multi-stage model shows how perceptual analyses, explicit classification, and cognitive mastering interact under the influence of prior experience, domain expertise, and personal taste. Early stages emphasize automatic, perceptual and affective processing, while later stages involve deliberate interpretation and evaluation. The proposed framework operationalizes the three components highlighted in red.}
\Description{Diagram of a multi-stage model of aesthetic experience in a museum or gallery context. The model shows a left-to-right processing flow from pre-classification of an artwork through perceptual analyses, implicit memory integration, explicit classification, and cognitive mastering, followed by evaluation. Perceptual analyses include visual features such as complexity, contrast, symmetry, order, and grouping. Implicit memory integration includes familiarity, prototypicality, and peak-shift effects. Explicit classification focuses on style and content. Cognitive mastering involves art-specific interpretation and self-related interpretation. Prior experience, domain-specific expertise, interests, and personal taste influence multiple stages. Emotional and affective states are continuously evaluated throughout the process. Aesthetic judgment emerges from the evaluation stage, and aesthetic emotion is described as a by-product of processing. Three stages—perceptual analyses, explicit classification, and cognitive mastering—are highlighted to indicate the components operationalized by the framework.}
\label{fig:leder_model}
\end{figure}

\subsection{Modeling Human Experience of Product Aesthetics}
Advances in the psychological and cognitive sciences provide valuable frameworks for understanding consumers' aesthetic experience with products. Crilly et al. proposed a communication model of design, which treats product form as signs that are interpreted by users, and describes a number of stages that exist between the \textit{source} (designers' expression of intent) and \textit{destination} (cognitive, affective, and behavioral response), including the \textit{transmitter} (visual aspects of the product, such as geometry and texture), \textit{channel}, and \textit{receiver} (physiological senses) stages~\cite{crilly2004seeing}. 

On the other hand, cognitive information processing models of aesthetic experience posit that individuals interpret and evaluate aesthetic objects through multiple cognitive stages, including perceptual processing (e.g., visual complexity, contrast) and explicit classification (e.g., genre of art, product category), among others, likely resulting in an emotional reaction and higher-order judgment (e.g., aesthetic preference, behavior)~\cite{reber2004processing,leder2004model}. Both models offer valuable perspectives that inform a holistic approach to understanding consumers' experience with products. In particular, in bridging subjective aesthetic experience with design intent as a communicative act, it is important to consider the link between perception and high-level judgment. 

For operationalizing these models, Kansei (or affective) design have been widely explored in computational creativity, particularly in the development of intelligent systems that support an emotional and sensory-driven design process by translating subjective human perceptions into concrete product parameters~\cite{nagamachi2002kansei}. Kansei design typically follows a structured process that consists of capturing emotional needs through user input (including online reviews and surveys), quantifying and analyzing these needs to map them to specific design attributes, and interpreting the results to refine product design for improved emotional resonance. Recent studies have leveraged natural language processing and deep learning approaches to map affective responses to design attributes, and even generating new designs in ways that improve personalization in various domains such as automotive design and consumer goods~\cite{chen2024autospark,yang2023product}. 

Recent work has also examined how individual differences, design attributes, and cultural factors jointly influence consumer preferences. Myszkowski and Storme find that personality traits such as openness to experience predict design-driven consumer decisions~\cite{Myszkowski2012}, while Silvia and Barona show that expertise moderates preference for curved versus angular forms~\cite{silvia2009people}. Fujiwara and Nagasawa highlight cultural specificity in brand aesthetics, showing that Japanese consumers associate luxury with particular psychological markers~\cite{fujiwara2015analysis}. While such approaches have been applied successfully to enhance the emotional appeal of consumer products, they do not consider designers' communicative intent and lack generalizable insights into why designs are perceived in a certain way.

\subsection{Integrating Human- and Computationally-Derived Features}
Recent work in human-computer interaction (HCI) and machine learning (ML) applications examined the use of low level features such as hue, contrast and texture to predict high-level semantic attributes of visual stimuli and derive higher-order judgments such as aesthetic preference and behavior. Shin et al. proposed a deep ML framework to transform design samples into low-dimensional feature vectors before extracting key representational design features to compute a utility function for each individual~\cite{shin2024learning}. This approach allowed them to use low-level visual cues to predict user preferences at both the individual and group levels (e.g., age, gender). Similarly, Iigaya et al. demonstrated that aesthetic preferences for art can be predicted from a mixture of low-level visual features and artistic expression (e.g., low: hue, saturation; high: dynamic), emphasizing the interplay between these features in shaping subjective aesthetic judgments~\cite{iigaya2021aesthetic}. Wu et al. investigated how user-perceived brand personality could be inferred from visual patterns of mobile application interfaces, showing that low-level UI elements like color and layout contribute to high-level brand perception~\cite{wu2019understanding}. Guo et al. explored how first impressions of images in online medical crowdfunding campaigns could be modeled, finding that a combination of low-level and domain-specific visual features such as facial expressions in promotional images significantly influence viewers' emotional and cognitive responses, which subsequently affect donation decisions~\cite{guo2022understanding}.

Together, these studies highlight the critical role of low-level (or computer vision proxies thereof) visual features and high-level constructs (e.g., emotion, brand personality) in bridging the gap between product appearance and higher-order subjective judgments. However, it is not clear how product designers can make use of, or integrate the identified features into their own domain-, brand-, and team-specific creative processes and language of practice. In this paper, we aim to utilize a combination of designer-informed, human-derived classifications of design features and machine-extracted proxies of low-level visual features to understand ways in which the features facilitate guided design decision making in the context of product design.
\section{Operationalizing the Model of Aesthetic Experiences}
In constructing the computational framework, we operationalize three components of Leder et al.'s~\cite{leder2004model} model of aesthetic experience by integrating machine-extracted perceptual features, designer-informed domain-specific design patterns, and computational measures of consumers' self-related style interpretations within the wheel design domain.
\subsection{Machine-Extracted Visual Features}
We treat computer-vision-based visual features as proxies for the human perceptual system ("Perceptual Analyses" component in Figure~\ref{fig:leder_model}). In the wheel design domain, we identified color, texture profiles, and orientation as key features capable of capturing complex and fine-grained visual differences among designs. Although prior work~\cite{wu2019understanding} has examined perceptual features such as symmetry, order, and grouping, these dimensions are either uninformative for wheel spoke configurations, which are inherently radially symmetric, or remain difficult to operationalize using existing computer-vision techniques.

\subsection{Designer-Informed Domain-Specific Design Patterns}
To model design-domain knowledge, we focused on human-recognizable wheel design patterns informed by wheel product marketing language and expert input from professional wheel designers. Table~\ref{tab:designerfeatures} presents representative wheel images alongside associated pattern descriptions, such as double-spoke and offset-spoke configurations. We posit that incorporating these domain-specific design patterns (“Explicit Classification” component in Figure~\ref{fig:leder_model}) is essential for translating model outputs into interpretable insights that align with designers' existing vocabulary and design practices. We did not include the 'content' construct in our model since it did not apply to the wheel design domain.

\subsection{Consumer's Own Interpretation of Style}
According to Leder et al., naïve perceivers often interpret artworks by relating their content to personal situations and emotional states~\cite{leder2004model}. This process can elevate aesthetic evaluations when an object evokes familiar or meaningful visual associations. For example, Alessi kitchen products may be perceived as more aesthetically pleasing than conventional kitchen utensils because they deliberately exploit visual familiarity; the Alessi citrus juicer, for instance, resembles a calamari squid. However, the same object may elicit different familiar associations across individuals, such as a spider or an alien, leading to divergent interpretations. Martindale argued that the number and diversity of semantic associations activated by a stimulus reflect the depth of a viewer's understanding and self-related interpretation of an artifact~\cite{martindale1984pleasures}. Building on this perspective, we operationalized self-related interpretation (“Cognitive Mastering” component in Figure~\ref{fig:leder_model}) using a computational measure (inspired by~\cite{chen2024understanding}) of semantic distance between recognizable visual features of wheel images, generated by a vision language model, and participants' stated decision criteria for interpreting each style aesthetic, elicited by the product images. We hypothesized that smaller semantic distance would be positively related to aesthetic judgments.
\section{Method}
\begin{figure*}[t]  % 't' for top placement
\centering
\includegraphics[width=1\textwidth]{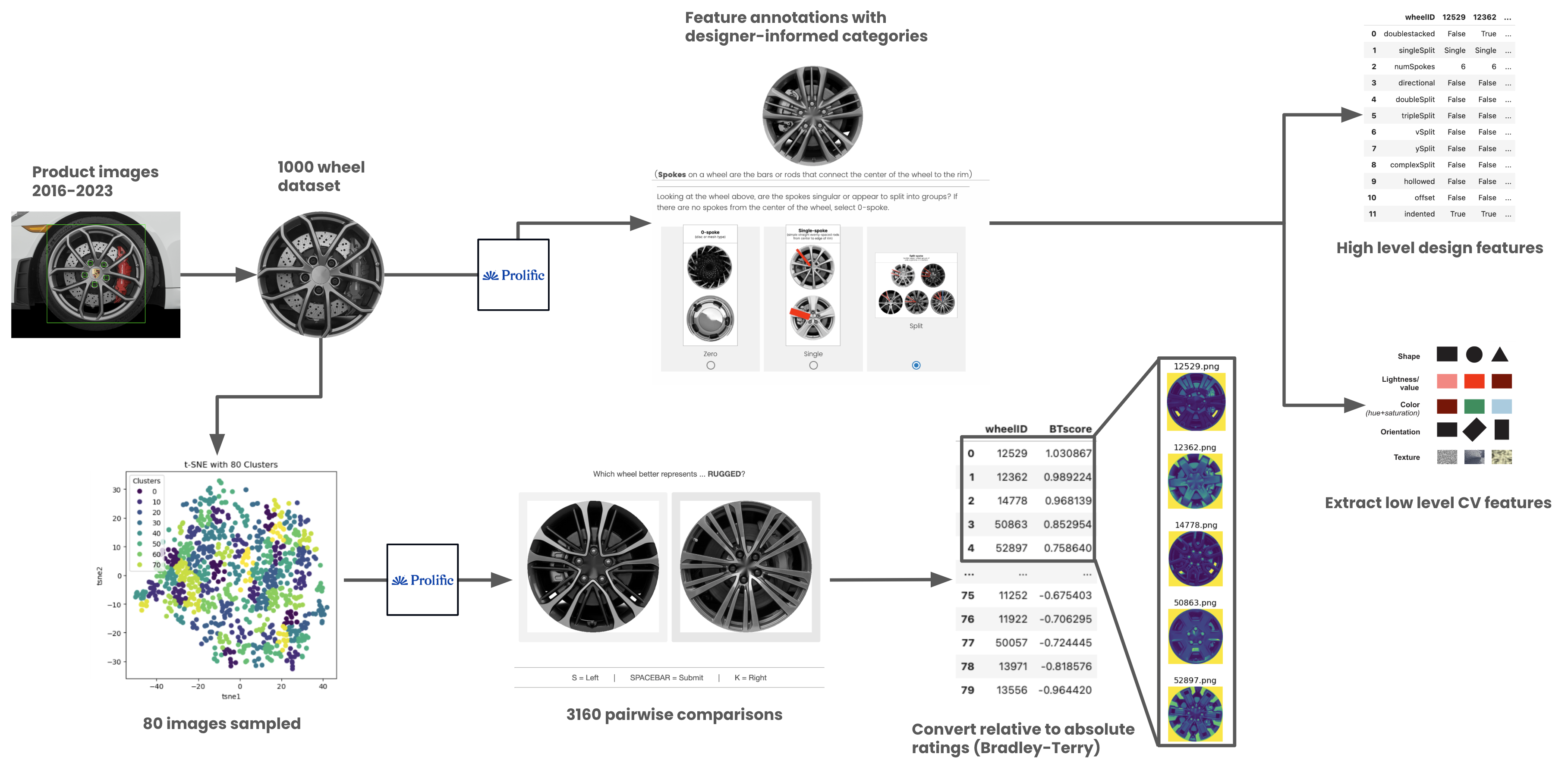} 
\caption{Data collection pipeline involving human subjects. Starting from 1000 curated wheels, we derived high-level visual features from crowd-based feature annotations informed by designer-informed classifications. We used a subset of the wheel dataset to run multiple pairwise comparisons with human raters, which enabled subsequent analyses after converting pairwise ratings into Bradley-Terry scores.}
\Description{Diagram of a data collection and analysis pipeline for automotive wheel design images. The pipeline starts with product images of wheels from 2016 to 2023, forming a curated dataset of 1,000 wheel images. These images are used in two parallel processes. In the first, crowd workers annotate visual features of wheels using designer-informed categories such as shape, spoke count, color, orientation, and texture, producing high-level design features and corresponding low-level computer vision feature maps. In the second, a subset of 80 sampled images is visualized using a t-SNE plot and used to generate 3,160 pairwise comparisons where human raters judge relative design qualities between wheel images. The pairwise comparison results are converted into absolute scores using a Bradley–Terry model. Arrows indicate data flow between stages, showing how annotations and comparisons support downstream quantitative analysis of wheel aesthetics.}
\label{fig:datatwocolumn}
\end{figure*}

\subsection{Data Collection}
We use car wheels as a case study to demonstrate the application of our integrated framework. There are two reasons for doing so. First, although the design space of car wheels is limited to a circular object, the diverse design choices available in today’s market enable us to analyze how various features influence consumer perception. Car exteriors, on the other hand, contain significantly less diversity due to performance constraints and safety regulation. Second, while wheels play an important aesthetic role, they are less integrated into a car's overall design than elements such as the front grille or tail lamps. This relative independence from engineering and manufacturing constraints allows us to isolate their visual impact with minimal interference from other stylistic features. In order to address the aforementioned research questions, we used both computer vision techniques and surveyed human participants to extract features meaningful to design practice.

%Since wheels can be swapped or customized, they provide an ideal opportunity for studying consumer preferences in a controlled setting.

\subsubsection{Styling Keywords and Designer-Informed Visual Features}
%, we drew on a larger, ongoing study informed by prior work on designer–AI collaboration~\cite{jeon2024weaving}.
To better align with the context of product design and development, we consulted six practicing subject matter experts\footnote{This is a small team of professional exterior designers in a US-based design studio that is focused on designing automotive products in the North American market.} in a large automotive manufacturing company. Drawing from prior work~\cite{jeon2024weaving} on designer–AI collaboration that identified an initial set of 25 wheel style keywords, we asked designers about styling keywords used in practice, from which we then sub-sampled nine keywords that informed subsequent data collection. The nine style keywords included: Aerodynamic, Classic, Dynamic, Elegant, Futuristic, Luxury, Rugged, Sleek, Sporty. Through further discussions, and our assessment of labels in the market used to describe different wheel design features, we identified four levels of distinction---directionality, split type, spoke count, and arrangement of spokes (Appendix \ref{appx:wheelfeatures})---that help organize the designer-informed wheel design features. 

\subsubsection{Image Stimuli Preparation}
We used 1000 wheel product images\footnote{The images used in this study are the property of EVOX Productions, LLC and are subject to copyright law. The appropriate licenses and permissions have been obtained to ensure the rightful use of these images in our study. For more information, visit \url{https://www.evoxstock.com/}.} across all car brands available in the US market between 2016 and 2023 to prepare our image stimuli. The stimuli were used to elicit human annotations of wheel design features, pairwise comparisons against high-level attributes, and computer vision-based extraction of low-level visual features. After downloading and removing duplicate images, we applied computational image editing techniques (e.g., mask, blur) to eliminate potential biases arising from brand awareness. We trained a YOLOv9 classifier on the CAWDEC dataset~\cite{stanek2023realtime} to detect the wheel center and rim positions. After creating a circular mask, we used inpainting to remove logos displayed on the center hub cap and blur it with the rest of the wheel. To further conceal branding information from the brake calipers and edge of rims, we first created a mask and then programmatically applied a blur effect to the targeted areas. Finally, all images were converted to grayscale to ensure visual consistency and reduce any influence of vivid colors\textendash{}such as those found on certain sports car brake calipers. This process yielded a final dataset of 1000 wheel images.

\subsubsection{Sampling Representative Designs}
To reduce the number of image stimuli utilized in the pairwise comparisons task, we downsampled our dataset to 80 images. To do this, we first used a Vision Transformer (ViT)~\cite{dosovitskiy2020image} model to extract representative visual features from the set of 1000 wheel images and reduced the data to two dimensions with a t-SNE algorithm~\cite{van2008visualizing}. After applying k-means clustering, we evaluated the distances between the cluster centroids and all data samples in each case and selected the 80 samples that were closest to their respective centroids across all cases.
%We then used the Elbow method~\cite{marutho2018determination} to determine an appropriate number of clusters and
\begin{figure*}[t]
\centering
\begin{minipage}[t]{0.45\textwidth}
  \centering
  \includegraphics[width=\linewidth]{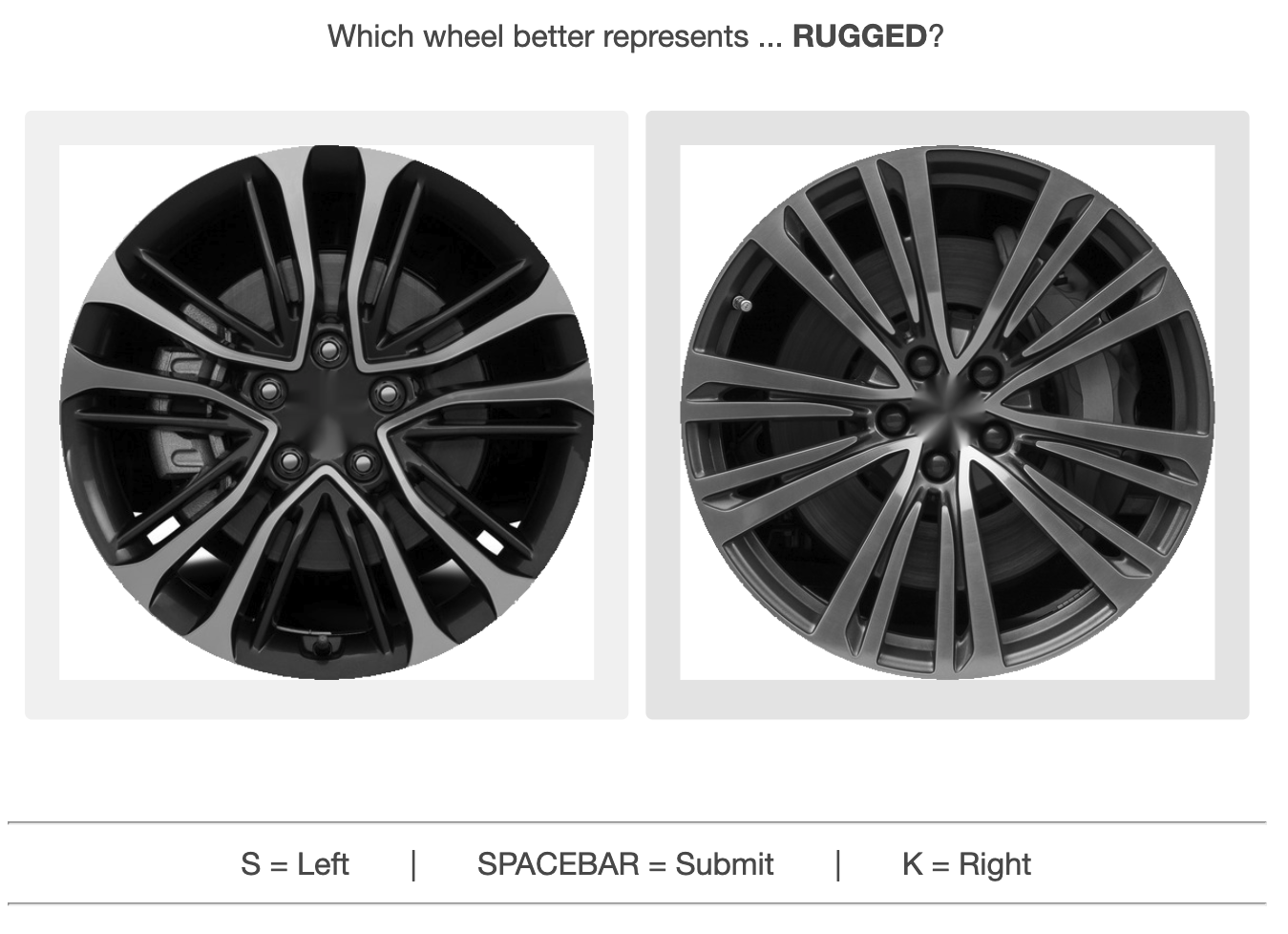}
      \captionof{figure}{Pairwise comparison task. Participants selected one of two wheels, with each viewing a unique set of 240 wheel pairs randomly sampled from the 3,160 possible pairwise combinations.}
      \Description{Interface for a pairwise comparison task asking which wheel better represents the attribute “rugged.” Two wheel images are shown side by side on a neutral background, one on the left and one on the right. Participants select either the left or right wheel using keyboard inputs, with on-screen instructions indicating key mappings for selection and submission.}
  \label{fig:pairwise_comp}
\end{minipage}
\hfill
\begin{minipage}[t]{0.45\textwidth}
  \centering
  \includegraphics[width=\linewidth]{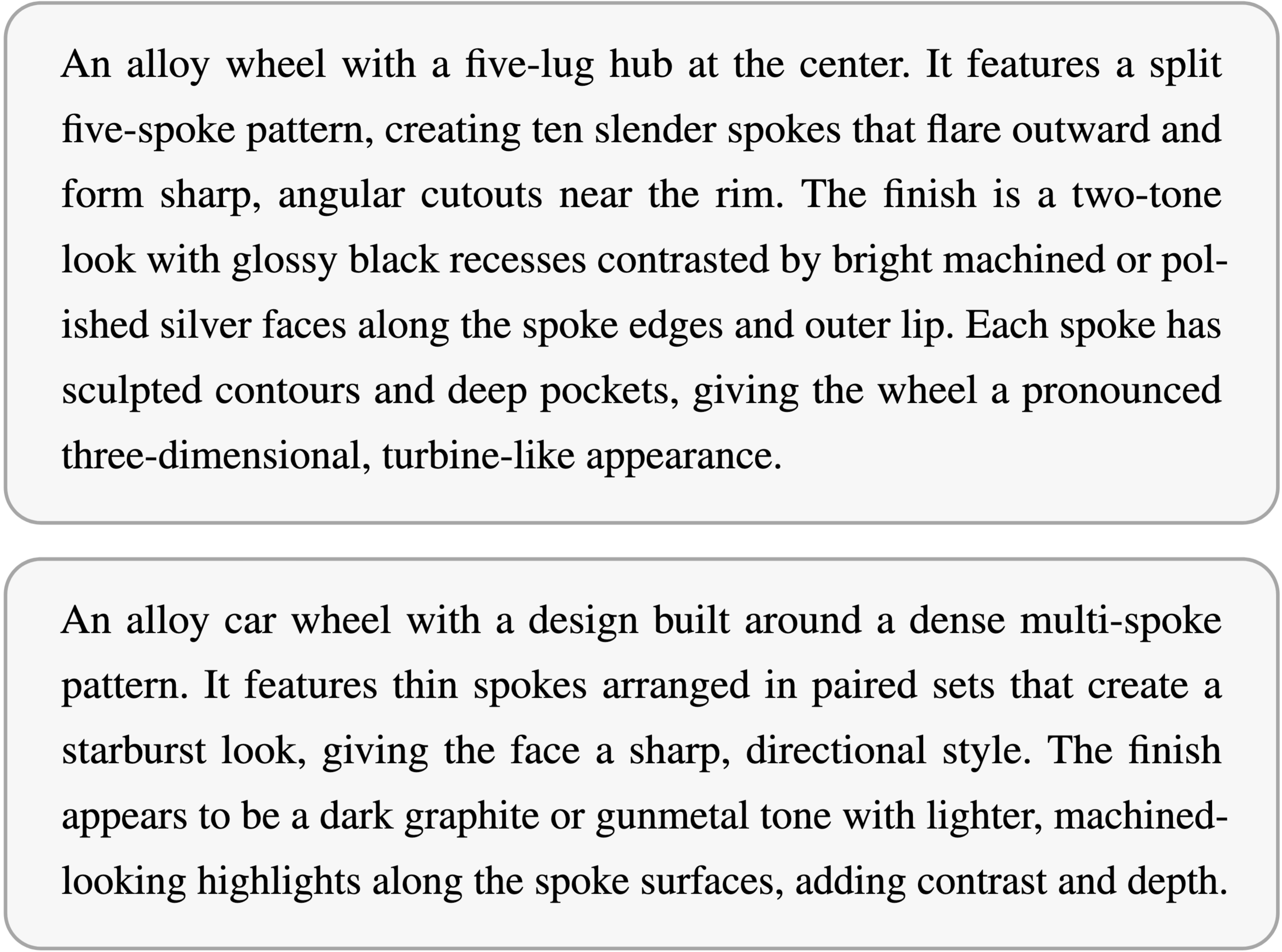}
  \captionof{figure}{Wheel design image captions generated by OpenAI GPT-5.2. The top and bottom captions describe left and right wheel images in Figure \ref{fig:pairwise_comp}.}
  \Description{Two text captions describing the left and right wheel images from the pairwise comparison task. The top caption describes a split five-spoke alloy wheel with a glossy black and polished silver finish and a sculpted, turbine-like appearance. The bottom caption describes a dense multi-spoke wheel with paired thin spokes and a dark graphite finish with lighter highlights. The captions are presented as model-generated textual descriptions of the corresponding wheel images.}
  \label{fig:gpt5-output}
\end{minipage}
\end{figure*}
%\begin{figure}[t]  % 't' for top placement
%    \centering
%    \includegraphics[width=0.7\linewidth]{fig/appx_pairwise_comp.png}
%    \caption{Pairwise comparisons task.}
%    \label{fig:pairwise_comp}
%\end{figure}

%\begin{figure}[t]
%\centering
%\begin{tcolorbox}[llmoutput,width=0.5\linewidth]
%{\footnotesize
%An alloy wheel with a five-lug hub at the center. It features a split five-spoke pattern, creating ten slender spokes that flare outward and form sharp, angular cutouts near the rim. The finish is a two-tone look with glossy black recesses contrasted by bright machined or polished silver faces along the spoke edges and outer lip. Each spoke has sculpted contours and deep pockets, giving the wheel a pronounced three-dimensional, turbine-like appearance.
%}
%\end{tcolorbox}
%\begin{tcolorbox}[llmoutput,width=0.5\linewidth]
%{\footnotesize
%An alloy car wheel with a design built around a dense multi-spoke pattern. It features thin spokes arranged in paired sets that create a starburst look, giving the face a sharp, directional style. The finish appears to be a dark graphite or gunmetal tone with lighter, machined-looking highlights along the spoke surfaces, adding contrast and depth.
%}
%\end{tcolorbox}
%\caption{Wheel design image captions generated by OpenAI GPT--5.2. The top and bottom captions describe left and right wheel images in Figure \ref{fig:pairwise_comp}.}
%\label{fig:gpt5-output}
%\end{figure}

\subsubsection{Pairwise Comparisons}\label{sec:pairwisecomparisons}
Through prior discussions with automotive designers, we identified a set of nine style keywords that are commonly used in guiding the wheel design process---including \textit{classic}, \textit{sporty}, \textit{luxurious}, \textit{futuristic}, \textit{rugged}, \textit{dynamic}, \textit{aerodynamic}, \textit{sleek}, and \textit{elegant}. To obtain human aesthetic ratings of wheel images we collected pairwise comparisons from the crowd via Prolific\footnote{\url{https://www.prolific.com/}} where we asked each participant to choose only one of the two randomly presented wheels that best matched the aesthetic style evoked by a given keyword such as \textit{rugged} (Figure \ref{fig:pairwise_comp}). Out of the 80 representative wheels, there were 3,160 possible unique pairwise combinations. Each participant viewed 240 randomly sampled pairs presented through the Qualtrics\footnote{\url{https://www.qualtrics.com/}} survey platform. In total, this setup yielded 575,520 ratings across the nine style keywords from 2,398 participants, with each of the 3,160 pairs receiving between 10 and 41 ratings and an average of 20.24 ratings.

To prepare for analysis, we fit a Bradley-Terry model~\cite{bradley1952rank} to estimate latent preference scores for each stimulus based on pairwise comparisons. These scores represent relative importance, where a higher score indicates a greater likelihood of being selected over other items in the set. A wheel with a higher Bradley-Terry score indicates that participants are more consistently linking it to a specific style keyword.
%to convert the relative ratings into absolute scores, where each stimulus is assigned a numerical value representing its overall importance or preference compared to other items in the set.

At the end of the pairwise comparisons task, participants were asked to provide free text responses to the question: ``\textit{List any visual features (what you saw in the images) that you used as criteria to choose the more [style keyword] wheel}.''

\begin{table*}[t]
    \centering
    \footnotesize
    \renewcommand{\arraystretch}{1} % Increase row height for readability
    \setlength{\tabcolsep}{5pt} % Adjust column separation
    \begin{tabular}{p{1cm} p{3cm} p{10cm}}
        \textbf{Type} & \textbf{Features} & \textbf{Description}\\
        \toprule
        Color & \ms{value} & Mean brightness or intensity of a color, ranging from completely dark to fully bright\\
        \midrule
        Texture & \ms{keypoints} & Number of distinctive local texture features (i.e., key points) in images\\
        & \ms{tamura.coarseness} & Granularity of a texture; high values indicate large, distinct patterns or structures\\
        & \ms{tamura.contrast} & Overall intensity distribution across the image; high values indicate strong differences between light and dark areas (e.g., sharp shadows).\\
        & \ms{tamura.directionality} & Presence of oriented structures or patterns in the texture; high values indicating dominant orientation (e.g., striped fabrics).\\
        & \ms{glcm.contrast} & Intensity difference between neighboring pixels; high values indicate sharp transitions.\\
        & \ms{glcm.correlation} & Linear dependency between neighboring pixels; high values indicate a strong linear relationship.\\
        & \ms{glcm.energy} & Uniformity of the texture; high values indicate a more consistent texture.\\
        & \ms{glcm.homogeneity} & Closeness of the GLCM distribution to the diagonal; high values indicate a more uniform texture.\\
        \midrule
        Orientation & \ms{angles} & Number of distinct and dominant line orientations.\\
        \bottomrule
    \end{tabular}
    \caption{Overview of computationally derived low-level visual features, organized by type and individual features.}
    \Description{Table listing computationally derived low-level visual features organized by type, feature name, and description. Feature types include color, texture, and orientation. Texture features include keypoint counts, Tamura measures such as coarseness, contrast, and directionality, and gray-level co-occurrence matrix statistics such as contrast, correlation, energy, and homogeneity. Orientation features describe the number of dominant line angles in an image.}
    \label{tab:cvfeatures}
\end{table*}

\subsubsection{Feature Annotations}
Taking guidance from designer-informed categories (Table \ref{tab:designerfeatures}), we collected crowd annotations of wheel design features from 315 participants who each provided annotations for 20 wheels. In particular, we collected information about the number of spokes, split type (none, single, split), and presence and absence of design features in the form of TRUE or FALSE responses with each feature receiving at least 5 responses (ranging from 5 to 10 responses for each wheel). This yielded 6,300 responses in total.

\subsection{Computationally-Derived Visual Features}
\subsubsection{Color}
Since the wheels were displayed in grayscale, the only available color information was their brightness (value). To quantify this, we extracted the value channel from the HSV color space and calculated the mean brightness.

\subsubsection{Texture}
We extracted texture features using three different computer vision algorithms, including Scale Invariant Feature Transform (SIFT)~\cite{lindeberg2012scale}, Tamura texture features~\cite{tamura1978textural}, and Gray Level Co-occurrence Matrix (GLCM)~\cite{haralick1973textural}, following those similarly used in~\cite{machajdik2010affective}. Each feature extraction technique offers a slightly different interpretation of surface texture. Because each technique captures distinct and complementary aspects of surface texture, using multiple feature extraction approaches ensures a more robust and comprehensive interpretation of texture. SIFT detects and describes distinctive local texture features (i.e., key points) in images. Tamura features are based on human visual perception, and are specifically designed to capture how humans perceive texture qualities such as coarseness, contrast, directionality, and roughness. GLCM describes the variation in intensity in neighboring pixels in an image, and include contrast, correlation, homogeneity, and energy. Table \ref{tab:cvfeatures} shows the full list of low-level visual features we computed.  

\begin{figure}[b]
    \centering
    \includegraphics[width=0.4\linewidth]{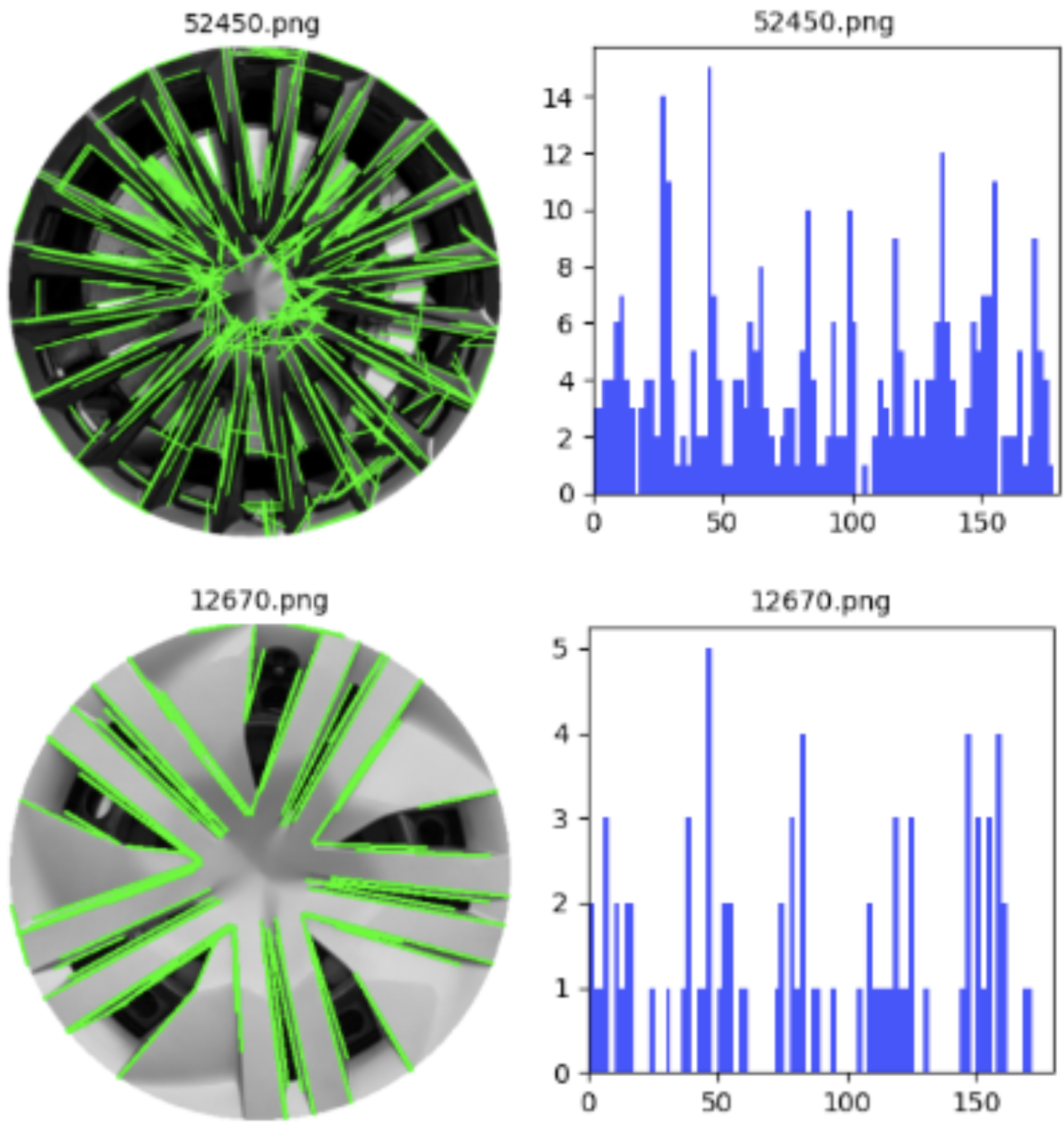}
    \caption{Visualization of dominant line orientations in a wheel image. The histograms show the frequency distribution of line segments counted in each angle bin.}
    \Description{Visualization of dominant line orientations for two wheel images. Each wheel image is overlaid with detected line segments shown in green, highlighting spoke directions. To the right of each image, a histogram shows the frequency of line segments across angle bins, illustrating differences in orientation distributions between a dense multi-spoke wheel and a simpler spoke design.}
    \label{fig:method_angles}
\end{figure}

\subsubsection{Orientation}
To determine the dominant line orientations on each wheel, we first applied Canny Edge Detection (with suitably chosen lower and upper thresholds) to highlight prominent edges~\cite{canny1986computational}. Next, we used Probabilistic Hough Transform (implemented via the \texttt{cv2.HoughLinesP} function in OpenCV) to detect individual line segments from the binary edge map~\cite{opencv_library,kiryati1991probabilistic} (Fig \ref{fig:method_angles}). The transform parameters---such as the detection threshold, minimum line length, and maximum gap between points---were empirically tuned to balance sensitivity and specificity.

For each detected line segment, we calculated its angle \(\theta\) in degrees using
\[
\theta = \mathrm{atan2}(\Delta y, \Delta x),
\]
and then normalized \(\theta\) to the range \([0,180)\) because orientations separated by 180\(^\circ\) represent the same physical direction. We grouped these angles into bins with 2\(^\circ\) increments to identify distinct families of orientations. Bins with at least two line segments were labeled as dominant orientations for the wheel.

\subsubsection{Image Captions}\label{sec:imagecaptions}
To complement computer vision and human-annotation features, we generated detailed text descriptions of 80 wheel designs using OpenAI's GPT-5.2 model\footnote{\url{https://openai.com/gpt-5/}}, which offers state-of-the-art visual perception and understanding performance. Each wheel image was processed through the model with a structured prompt requesting comprehensive descriptions of visual features, including spoke pattern, finish type, color scheme, geometric features, and stylistic attributes (see Appendix \ref{appx:prompt}). The resulting captions were standardized in both format and length, with each description containing approximately 100--150 words. These AI-generated descriptions served as the semantic representation of each wheel's visual properties for subsequent similarity analyses. Figure~\ref{fig:gpt5-output} shows example excerpts of the generated wheel descriptions.

\subsection{Analysis}\label{sec:analysis}
To answer RQ1, we treat the visual features (designer-informed, computationally-derived) as independent variables and Bradley-Terry (BT) scores as the dependent variable to predict the influence of visual features on consumer perceptions (BT scores) of wheel aesthetics against style keywords. For designer-informed features (Sec.~\ref{sec:rq1a}), we fit ordinary least squares (OLS) regression models predicting BT scores from visual feature presence. Binary design features were aggregated by computing the proportion of times each feature was annotated as present for a given wheel. For low-level computationally derived visual features (Sec.~\ref{sec:rq1b}), we similarly fit OLS regression models to estimate their associations with BT scores. All feature-based analyses were conducted separately for each of the nine style keywords to allow style-specific effects to emerge. 

Finally, we examine the semantic alignment between consumers' interpretations (Sec.~\ref{sec:rq1c}) and aesthetic perceptions. We first encoded both the LLM-generated wheel captions (Sec.~\ref{sec:imagecaptions}) and participant descriptions of visual features from the pairwise task (Sec.~\ref{sec:pairwisecomparisons}) for each style keyword into embedding vectors using Sentence-BERT \cite{reimers-2019-sentence-bert}. For each wheelID--style pair, we calculated cosine similarity between the wheel's caption embedding and all survey response embeddings for that keyword, then computed the mean similarity score as the aggregate measure of semantic alignment.

%To examine how semantic alignment relates to perceived aesthetic preference, we conducted Bayesian hierarchical regression analyses predicting BT scores from cosine similarity between participant-generated text descriptions and LLM-generated image descriptions. Cosine similarity was treated as a semantic alignment measure derived from participants' linguistic articulations of visual perception, rather than as a physical property of the stimuli. It was analyzed separately from regressions involving explicit design features, which represent structural attributes of the wheels.

To understand how semantic alignment relates to perceived aesthetic preference, we conducted OLS regression analyses predicting BT scores from cosine similarity between participant-generated text descriptions and LLM-generated image descriptions. Cosine similarity was treated as a semantic alignment measure derived from participants' linguistic articulations of visual perception, rather than as a physical property of the stimuli. It was analyzed separately from regressions involving explicit design features, which represent structural attributes of the wheels.

%BT scores were modeled as a linear function of cosine similarity, with style-specific intercepts and slopes estimated using partial pooling: \begin{equation}
%\text{BT}_i = \alpha_{s[i]} + \beta_{s[i]} \cdot \text{cos\_sim}_i + \epsilon_i
%\end{equation}
%where \textit{s}[\textit{i}] indexes the style keyword associated with observation \textit{i}. Style-specific intercepts ($\alpha_s$) and slopes ($\beta_s$) were modeled as draws from group-level normal distributions, allowing relationships to vary by style while borrowing statistical strength across styles. This hierarchical structure stabilizes estimates by shrinking noisy style-specific effects toward the population mean when evidence is weak.

%Residual errors were assumed to be normally distributed. Weakly informative priors were placed on all fixed effects (normal priors centered at zero), group-level standard deviations (half-normal or exponential priors), and the residual variance. Models were fit to 720 observations using the \verb|brms| package in R with four Markov chain Monte Carlo chains; all models showed satisfactory convergence ($\hat{R}\  \approx$ 1.0; effective sample sizes > 1000).
 %  brms formula: bt_score ~ 1 + cos_sim + (1 + cos_sim | style) %

To answer RQ2, we examine the spread of BT scores to assess variability across people's responses. For example, style keywords that have relatively large differences between the highest and lowest BT scores indicate that people are consistently linking specific wheel images with certain styles. Styles with small differences in BT scores indicate disagreement among participants, which could arise in one of the two ways: (1) high randomness in responses, where the identified visual design features are only weakly or inconsistently associated with the style keyword (i.e., responses are near chance level), or (2) high heterogeneity, where distinct subgroups of participants interpret the same style keyword differently, for example associating “aerodynamic” with either fewer or more open design elements.

We also examine variability in people's stated aesthetic perceptions elicited by styles (in response to viewing wheels) by analyzing the distributions of cosine similarity scores, which measure semantic alignment between LLM-generated wheel captions (Sec.~\ref{sec:imagecaptions}) and participant descriptions across the nine style keywords. In both analyses, we use Hartigan’s dip test \cite{hartigan1985dip} to assess the presence of unimodality in the response distributions.

%To do this, we generated detailed text descriptions of 80 wheel designs using OpenAI GPT-5.2 model. Each wheel image was processed through the model with a structured prompt requesting comprehensive descriptions of visual features including spoke pattern, finish type, color scheme, geometric features, and stylistic attributes. These AI-generated descriptions served as the semantic representation of each wheel's visual properties for subsequent similarity analyses.
\begin{figure}[t]  % 't' for top placement
\centering
\begin{subfigure}{0.5\columnwidth}
    \centering
    \includegraphics[width=\linewidth]{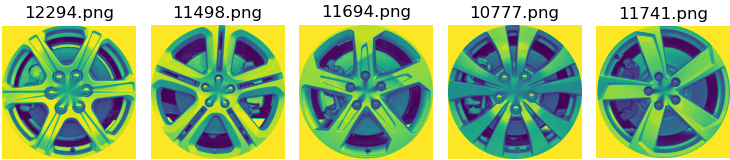}
    \caption{Classic}
    \label{fig:results_classic}
\end{subfigure}
\hfill
\begin{subfigure}{0.5\columnwidth}
    \centering
    \includegraphics[width=\linewidth]{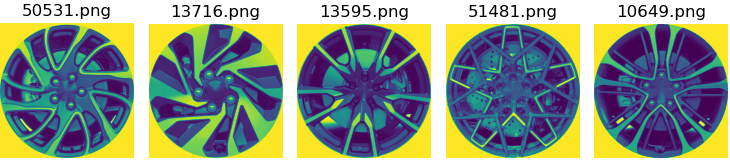}
    \caption{Dynamic}
    \label{fig:results_dynamic}
\end{subfigure}
\vspace{0.5em}
\begin{subfigure}{0.5\columnwidth}
    \centering
    \includegraphics[width=\linewidth]{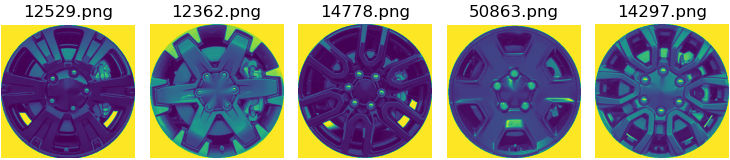}
    \caption{Rugged}
    \label{fig:results_rugged}
\end{subfigure}
\caption{Comparison of top five wheels for each style keyword rank-ordered by BT score. \textit{classic} styles are characterized by bright silver finishes and straight single or split spokes, whereas \textit{dynamic} styles feature a mix of thin, curved, and more complex spoke patterns with two-toned finishes. In contrast, \textit{rugged} styles generally include wheels with dark, thick spokes.}
\Description{Grid of wheel images showing the top five ranked examples for three style keywords: classic, dynamic, and rugged. Classic wheels display bright silver finishes and straight or split spokes. Dynamic wheels show thinner, curved, and more complex spoke patterns with higher visual variation. Rugged wheels feature darker finishes and thicker, heavier-looking spokes. Images are arranged in rows by style category.}
\label{fig:results_wheelscombined}
\end{figure}

\section{Results}
\subsection{RQ1. What Features Contribute Most to People's Perception of Aesthetic Style?}
\subsubsection{Influence of Designer-Informed Visual Features on Aesthetic Perceptions}\label{sec:rq1a}
\begin{figure*}[t]  % 't' for top placement
\centering
\begin{subfigure}{\textwidth}
    \centering
    \includegraphics[width=\textwidth]{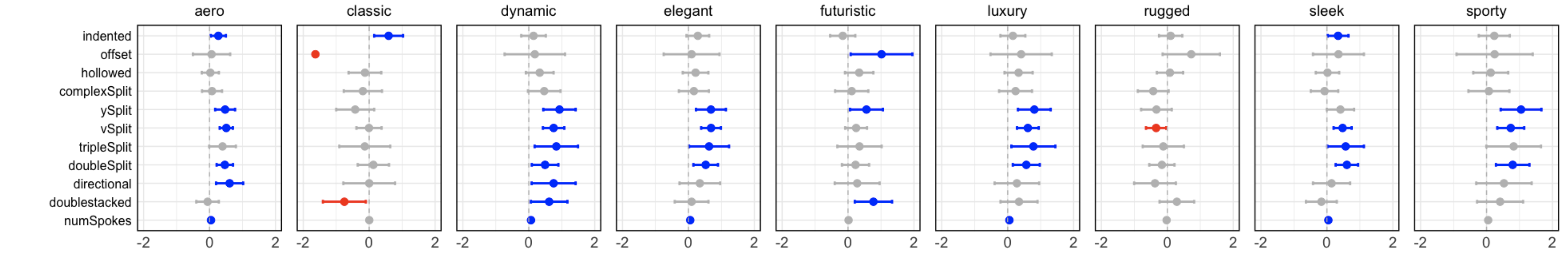}
    \caption{Influence of product features on aesthetic perceptions.}
    \label{fig:results_features}
\end{subfigure}
\vspace{1em}  % Add vertical space between subfigures
\begin{subfigure}{\textwidth}
    \centering
    \includegraphics[width=\textwidth]{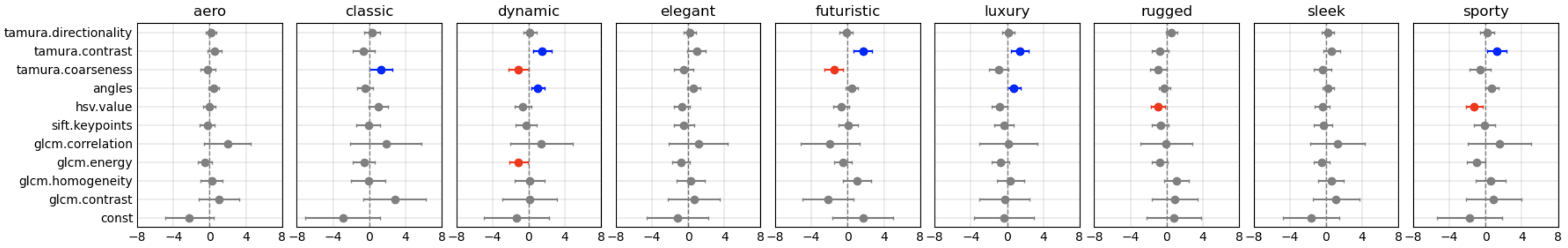}
    \caption{Influence of low-level visual features on aesthetic perceptions.}
    \label{fig:results_cvfeatures}
\end{subfigure}
\caption{Comparison of the influence of different visual features on aesthetic perceptions. Points correspond to linear regression beta weights for each feature and error bars indicate 95\% confidence intervals. Blue, red, and grey points indicate significantly positive ($p<.05$), significantly negative, and non-significant ($p>.05$) beta weights respectively.}
\Description{Two-panel figure comparing the influence of visual features on aesthetic perceptions across style keywords. The top panel shows regression coefficients for high-level product features such as spoke patterns and offsets, and the bottom panel shows coefficients for low-level visual features such as texture, contrast, and orientation. Points represent beta weights with horizontal error bars for 95 percent confidence intervals. Blue points indicate significant positive effects, red points indicate significant negative effects, and grey points indicate non-significant effects.}
\label{fig:results_combined}
\end{figure*}

The results of the linear regression reveal significant effects of various wheel design features on consumers' aesthetic evaluations using styling keywords. Across multiple styling keywords, the presence of different types of split-spoke designs, directional, and indented features emerged as key predictors. In particular, the \ms{ysplit}, \ms{vsplit}, and \ms{doublesplit} designs showed a consistently strong effect, significantly increasing ratings for styles such as \textit{aerodynamic}, \textit{dynamic}, \textit{elegant}, \textit{luxury}, and \textit{sleek} (Fig.~\ref{fig:results_features}). The \ms{directional} feature was also associated with higher ratings in the \textit{aerodynamic} and \textit{dynamic} styles, reinforcing the idea that consumers perceive these designs as more forward-looking and performance-oriented. Conversely, more intricate and fragmented spoke structures, such as \ms{doublestacked} and \ms{offset}, had more variance across different aesthetic dimensions. For instance, the \ms{doublestacked} design showed a positive influence on \textit{dynamic} ($\beta = 0.618$, $p = 0.032$) and \textit{futuristic} ($\beta = 0.770$, $p = 0.009$) evaluations, and was negatively associated with \textit{classic} ($\beta = -0.753$, $p = 0.027$), suggesting that consumers may perceive this feature as deviating from a more traditional aesthetic style.

The perception of \textit{dynamic} wheels appeared to be most influenced by a complex interplay of multiple features. The results suggest that \ms{doublestacked} ($\beta = 0.618$, $p = 0.032$) and \ms{directional} ($\beta = 0.753$, $p = 0.030$) significantly contribute to a \textit{dynamic} aesthetic (Fig \ref{fig:results_dynamic}, first two wheels from right and left respectively). Additionally, intricate spoke patterns, including \ms{complexsplit} ($\beta = 0.468$, $p = 0.067$)---exhibiting slight positive correlation---and the remaining split spokes ($\beta \in [0.494, 0.931$], $p < 0.03$), further reinforce this \textit{dynamic} perception, suggesting that consumers associate visual complexity with this style keyword. On the contrary, \textit{classic} and \textit{rugged} wheels displayed an insignificant or inverse relationship with most of the same features that drive the perception of a \textit{dynamic} aesthetic. This result implies that consumer perceptions can be effectively captured by the presence or absence of a specific set of design features.

\begin{figure*}[t]  % 't' for top placement
\centering
\includegraphics[width=\linewidth]{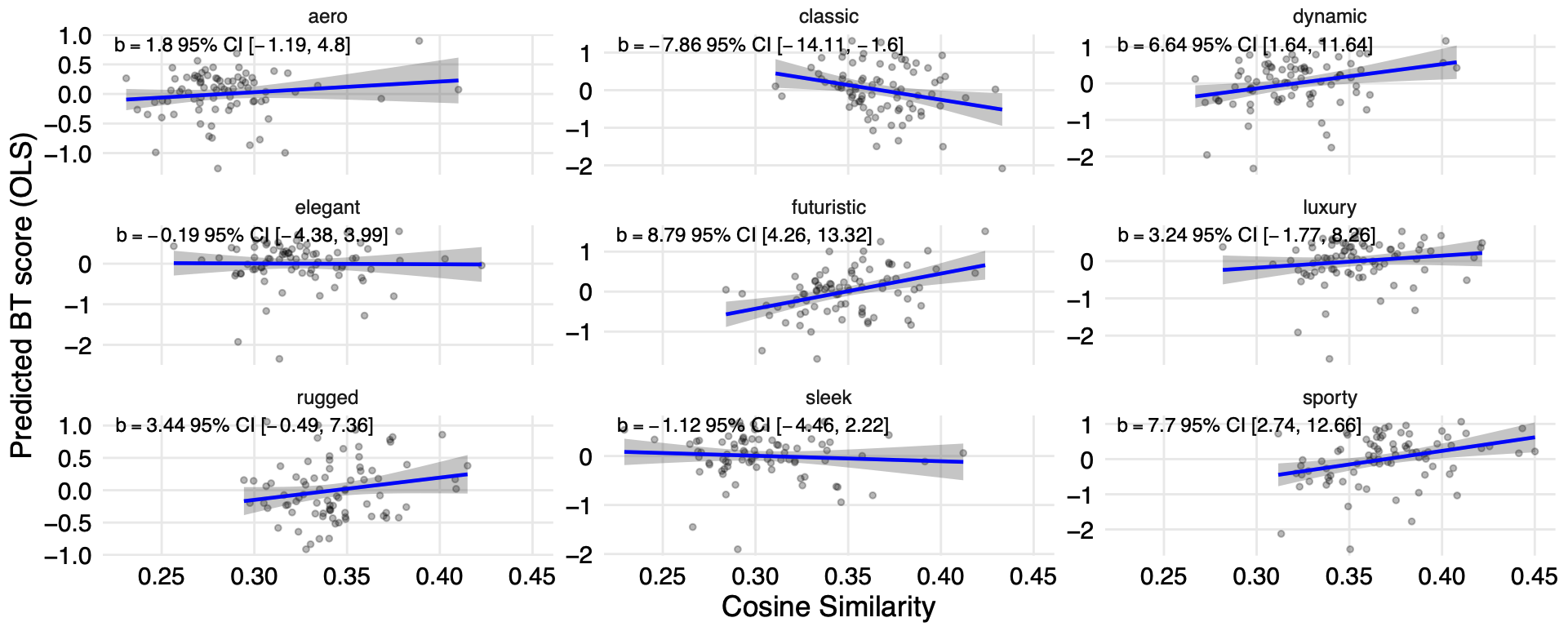} 
\caption{Relationships between cosine similarity and Bradley–Terry (BT) scores across nine style keywords. Each subplot shows individual wheel images as grey dots, with a regression line in blue, and the corresponding slope (b) and 95\% confidence intervals (95\% CI). \textit{dynamic}, \textit{futuristic}, and \textit{sporty} styles exhibit positive directional trends, whereas classic shows a negative trend, while others display weak associations. Results suggest style-dependent relationships between computational similarity measures and human perceptual judgments.}
\Description{Grid of nine scatter plots, one per style keyword, showing relationships between cosine similarity on the x-axis and predicted Bradley–Terry scores on the y-axis. Each subplot contains grey points representing individual wheel images, a blue regression line, and annotated slope estimates with 95 percent confidence intervals. Dynamic, futuristic, and sporty styles show positive trends, classic shows a negative trend, and other styles exhibit weak or near-zero associations.}
\label{fig:cosineBTregression}
\end{figure*}

\subsubsection{Influence of Computationally Derived Low-Level Visual Features on Aesthetic Perceptions}\label{sec:rq1b}
The results of our regression analysis indicate that certain low-level visual features of car wheels are predictive of consumer aesthetic evaluations, with varying degrees of influence across different aesthetic style dimensions. 

\ms{tamura.contrast} shows strong positive correlations with \textit{dynamic} ($\beta = 1.510$, $p = 0.007$), \textit{futuristic} ($\beta = 1.690$, $p < 0.001$), \textit{luxury} ($\beta = 1.430$, $p = 0.006$), and \textit{sporty} ($\beta = 1.270$, $p = 0.023$). \ms{angles}, a feature describing the number of distinct family of line orientation, is positively correlated with \textit{dynamic} ($\beta = 1.0222$, $p = 0.009$) and \textit{luxury} ($\beta = 0.772$, $p = 0.031$), and shows a non-significant trend with \textit{sporty} ($\beta = 0.7092$, $p = 0.069$). \ms{tamura.coarseness} showed strong positive correlation with \textit{classic} ($\beta = 1.279$, $p = 0.053$) but was negatively correlated with \textit{dynamic} ($\beta = -1.143$, $p = 0.048$), \textit{futuristic} ($\beta = -1.514$, $p = 0.006$) and, to a lesser degree, luxury ($\beta = -0.922$, $p = 0.083$). These results suggest that the mixed influence of high \ms{tamura.contrast} (greater differences between light and dark areas), high \ms{angles} (larger number of dominant line orientations), and low \ms{tamura.coarseness} (increased granularity of repeating textural patterns), may help describe the aesthetic style of \textit{dynamic} (Fig \ref{fig:results_dynamic}), \textit{futuristic}, \textit{luxury} and \textit{sporty} wheels. 

\ms{value}, a feature from the HSV color channel describing the brightness of wheels, is negatively correlated with \textit{rugged} ($\beta = -0.917$, $p = 0.023$) and \textit{sporty} ($\beta = -1.240$, $p = 0.011$), while slightly positively associated with \textit{classic} ($\beta = 1.030$, $p = 0.059$) (see Figure \ref{fig:results_wheelscombined}).

Altogether, these results show the complex interplay between the low-level visual features in consumer aesthetic judgments and highlight their potential in offering computational means of complementing consumer judgments. A prominent example is \textit{aerodynamic}, \textit{elegant} and \textit{rugged}, and \textit{sleek}, which show little to no statistically significant relationships with any machine-extracted visual features, yet could be described with human-derived visual features such as \ms{indented}, \ms{vsplit}, among others.

\subsubsection{Alignment Between Consumer's Aesthetic Interpretations and Perceptual Features}\label{sec:rq1c}
We conducted linear regression analyses to examine whether semantic alignment between LLM-generated descriptions of perceptual features and participants' stated features predicted BT scores across nine style keywords. In an additive model that also included cosine similarity and style word as predictors, cosine similarity showed a positive average association with BT scores; however, the model explained little variance overall (adjusted $R^2 \approx .006$), indicating limited explanatory power when effects are averaged across styles. We then fit an interaction model between cosine similarity and style word to test the effect of semantic alignment across style words. Allowing the slope of cosine similarity to vary by style significantly improved model fit relative to the additive model ($F(8,702) = 4.60$, $p < .001$), indicating that the relationship between semantic alignment and BT scores differs across styles. This result suggests that the relationship between semantic alignment and aesthetic preference is strongly style-dependent, motivating detailed examination to estimate style-specific effects.

Figure~\ref{fig:cosineBTregression} shows predictive relationships between cosine similarity (Sec. \ref{sec:analysis}) and BT scores for each style keyword. The predictor showed clear positive associations with \textit{futuristic}, \textit{sporty}, and \textit{dynamic} styles. \textit{futuristic} exhibited the strongest relationship ($\beta = 8.79$, CI $[4.26, 7.80]$), suggesting that higher cosine similarity scores were associated with substantially higher aesthetic perceptions. \textit{sporty} also showed a significant positive effect ($\beta = 7.70$, 95\% CI $[2.71, 12.66]$), followed by \textit{dynamic} ($\beta = 6.64$, 95\% CI $[1.64, 11.64]$)). Although these models explain a modest proportion of variance ($R^2 = .08-.16$), the consistency and direction of effects suggest that the predictor captures aesthetic perceptions for these styles.

In contrast, we observed a weak negative relationship for \textit{classic} ($\beta = -7.86$, 95\% CI $[-14.11, -1.60]$), suggesting that greater alignment with generic classic descriptors corresponded to lower perceived style strength. Other styles showed weak or negligible relationships between cosine similarity and BT scores, indicating that semantic alignment was not uniformly informative across all style keywords.

%Three style keywords exhibited strong positive relationships between semantic alignment and perceived preference: \textit{dynamic} ($\beta = 6.64$, 95\% CI $[1.64, 11.64]$), \textit{futuristic} ($\beta = 8.79$, 95\% CI $[4.26, 13.32]$), and \textit{sporty} ($\beta = 7.70$, 95\% CI $[2.71, 12.66]$). In contrast, \textit{classic} ($\beta = -7.86$, 95\% CI $[-14.11, -1.60]$) exhibited a negative directional trend, suggesting that greater alignment with generic classic descriptors corresponded to lower perceived style strength. Other styles showed weak or negligible relationships between cosine similarity and BT scores, indicating that semantic alignment was not uniformly informative across all style keywords.

Overall, these results indicate that semantic alignment serves as a strong, style-dependent predictor of aesthetic perception. Styles such as \textit{futuristic}, \textit{sporty}, and \textit{dynamic} appear to be associated with more standardized and linguistically codable visual semantics, for which alignment between participant language and model-generated descriptions reliably tracks perceived preference. In contrast, other styles show weaker or inconsistent relationships, underscoring that semantic alignment does not function as a universal predictor across aesthetic categories.

\begin{figure}[t]  % 't' for top placement
\centering
\includegraphics[width=0.7\linewidth]{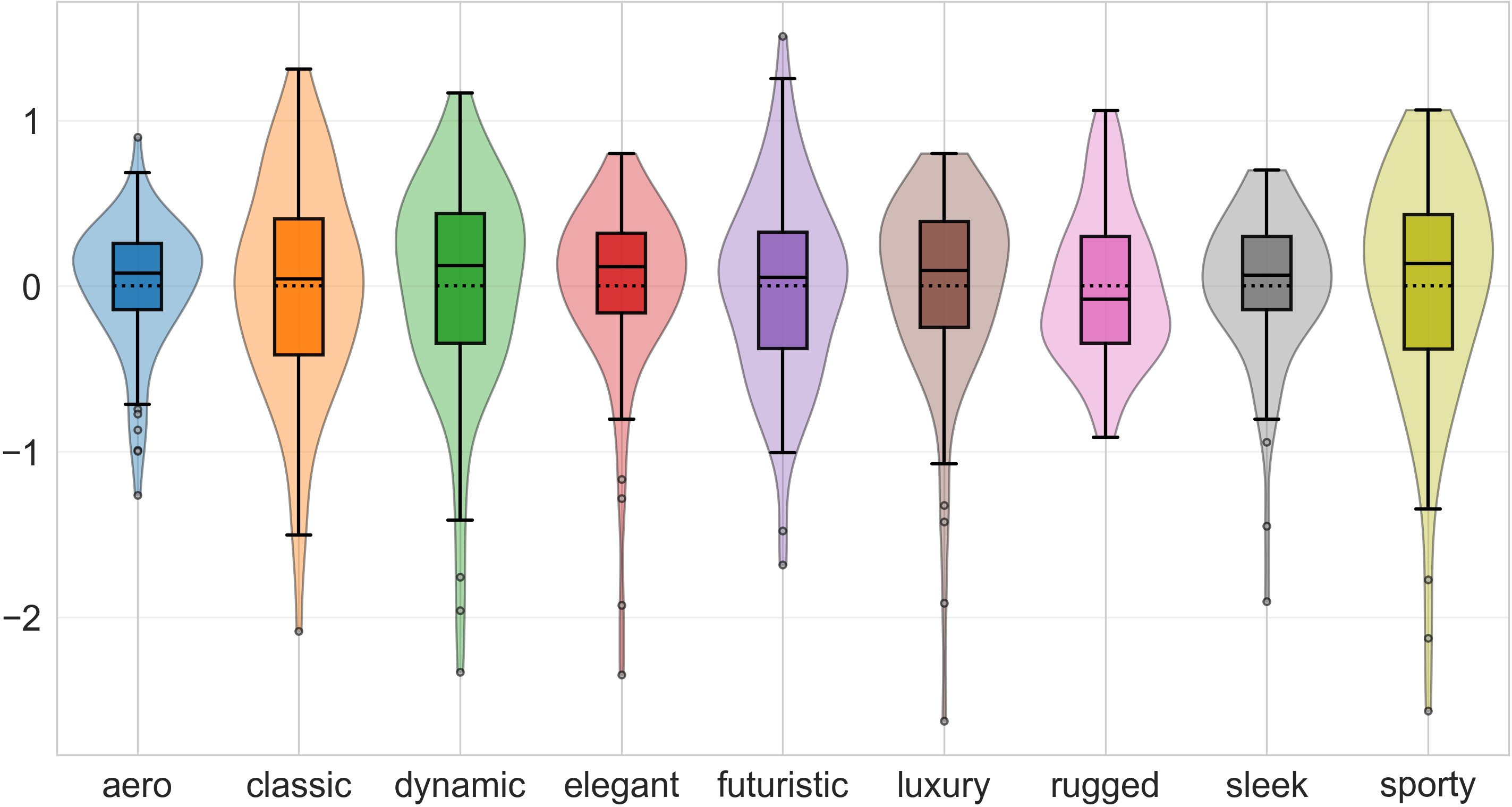}
\caption{Distributions of Bradley-Terry Scores (y-axis) by Style Keyword (x-axis). \textit{aero}, \textit{elegant}, and \textit{sleek} styles exhibit tighter interquartile ranges (IQR) compared to others. Styles such as \textit{classic} and \textit{sporty} show a wider spread, indicating that more participants favored one wheel over another in the middle range.}
\Description{Violin plots displaying distributions of Bradley–Terry scores across nine style keywords. Each plot includes a boxplot indicating median and interquartile range. Styles such as aero, elegant, and sleek show narrower distributions, while classic and sporty exhibit wider spreads, suggesting greater variability in participant preferences.}
\label{fig:results_BTstats}
\end{figure}

\subsection{RQ2. Do People Form Differentiable Perceptions of Aesthetic Style?}
\subsubsection{Differentiation in Aesthetic Judgments Across Styles}
We aimed to understand the extent to which people form consistent perceptions of product aesthetics and how much individual differences influence these perceptions. By examining the distribution of participants' ratings using Bradley-Terry scores, we report notable differences across various style keywords that reveal the level of differentiability in aesthetic evaluations (see Figure \ref{fig:results_BTstats}).

The standard deviations indicate variability in ratings, with \textit{classic} (0.656), \textit{sporty} (0.698), and \textit{dynamic} (0.654) showing the highest dispersion, implying there is a more pronounced difference between the most and least chosen wheels for each style. In contrast, \textit{rugged} (0.451), sleek (0.452) and \textit{aerodynamic} (0.388) showed lower standard deviations, suggesting reduced discriminability in human ratings.

The range of scores further highlights differences in judgment consistency. The lowest observed scores are notably negative across all attributes, with \textit{luxury} (-2.624) and \textit{sporty} (-2.563) experiencing the largest negative extremes. On the contrary, the highest scores range from \textit{sleek} (0.700) to \textit{futuristic} (1.509), showing that some images were perceived as highly dominant by certain participants. The interquartile range (IQR) provides further insights into rating variation. The middle 50\% of ratings (25-75th percentiles) are relatively narrow for \textit{rugged} (-0.344 to 0.299) and \textit{aerodynamic} (-0.142 to 0.259), indicating that participants did not overwhelmingly favor one wheel over another in the middle range. Meanwhile, \textit{classic} (-0.417 to 0.405) and \textit{dynamic} (-0.346 to 0.437) demonstrate wider IQRs, supporting the observation of greater variation. Hartigan's dip tests indicated no evidence of multimodality in BT score distributions across wheel designs for any style word (all $p > 0.45$), suggesting that BT scores vary continuously rather than forming discrete clusters.

\begin{figure}[t]  % 't' for top placement
\centering
\includegraphics[width=0.6\linewidth]{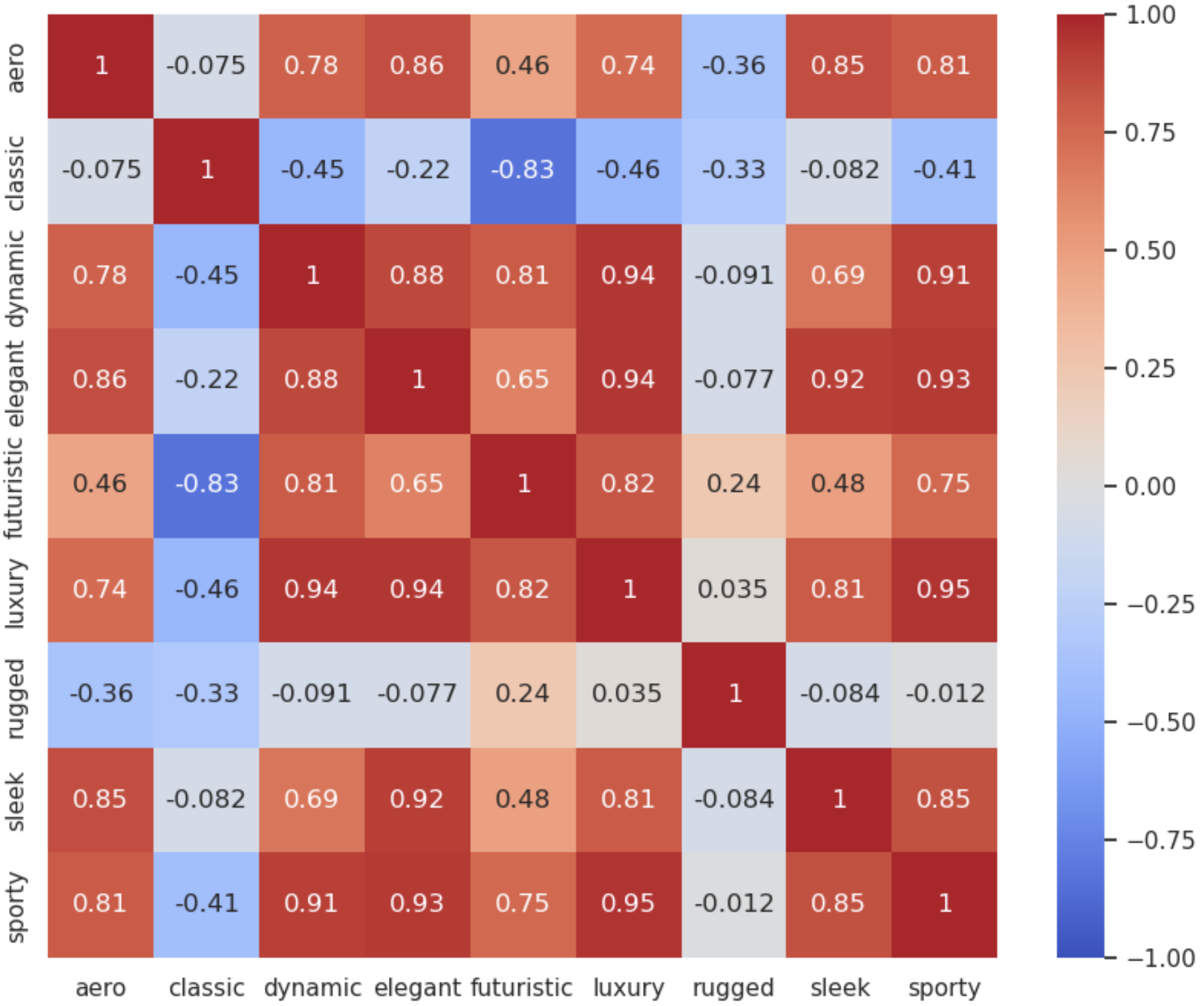} 
\caption{Correlation heat map of BT scores for style keywords. Strong positive correlations appear among styles such as dynamic, elegant, luxury, and sporty, while classic and rugged show weaker or negative correlations with several other styles.}
\Description{Heat map showing pairwise correlations between Bradley–Terry scores for nine style keywords. Rows and columns correspond to the same set of styles. Cell colors range from blue for negative correlations to red for positive correlations, with numeric values displayed in each cell. Strong positive correlations appear among styles such as dynamic, elegant, luxury, and sporty, while classic and rugged show weaker or negative correlations with several other styles.}
\label{fig:results_styleheatmap}
\end{figure}

The differences in variability in ratings could be explained by the correlations of BT scores for the style keywords. For instance, Figure \ref{fig:results_styleheatmap} shows a strong correlation across many style keywords such as \textit{luxury}, \textit{elegant}, \textit{sporty}, and \textit{dynamic}, which suggests that wheels perceived as luxurious are also likely to be perceived as \textit{elegant}, \textit{sporty}, etc. This correlation is also accentuated by the strong presence of a low level visual feature (i.e., \ms{tamura.contrast}) (Figure \ref{fig:results_cvfeatures}).

On the other hand, \textit{classic} and \textit{futuristic} displayed strong negative correlation, which was also exemplified via the striking contrast in the influence of product features, such as \ms{offset} and \ms{tamura.coarseness}, on aesthetic perceptions (Figure \ref{fig:results_features}). Finally, \textit{rugged} was weakly and slightly negatively correlated with most other styles, suggestive of its distinct style.

\begin{figure}[t]  % 't' for top placement
\centering
\includegraphics[width=0.7\linewidth]{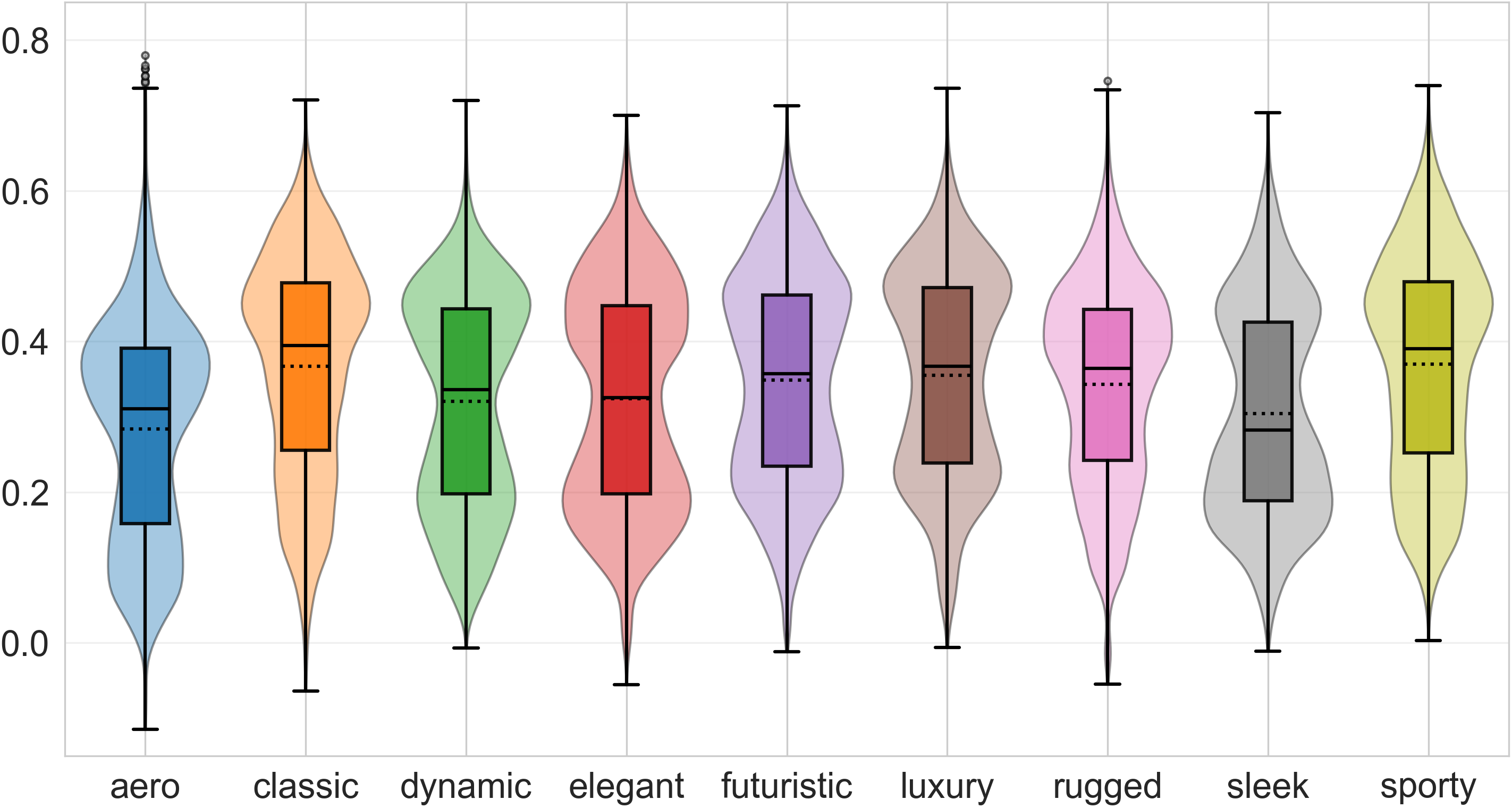}
\caption{Distributions of Cosine Similarity Scores (y-axis) by Style Keyword (x-axis). Cosine similarity scores are computed across text descriptions of wheel visual features generated by GPT-5.2 and those of survey respondents.}
\Description{Violin plots showing distributions of cosine similarity scores for nine style keywords: aero, classic, dynamic, elegant, futuristic, luxury, rugged, sleek, and sporty. Each violin includes an embedded boxplot indicating median and interquartile range. Distributions vary by style, with some showing tighter clustering and others broader spread, indicating differences in agreement between GPT-generated text descriptions and human survey descriptions.}
\label{fig:results_cosine}
\end{figure}
\subsubsection{Variation in Semantic Alignment Across Aesthetic Styles}
To understand why semantic alignment between language models and human interpretations is stronger for some styles than others, we examined the distribution of cosine similarity scores between LLM-generated captions and participant survey responses for each style keyword. Figure~\ref{fig:results_cosine} illustrates systematic differences in alignment strength across styles. The classic, rugged, and sporty categories exhibit higher median cosine similarity scores than other styles, indicating stronger alignment between the language model's descriptions and participants' interpretations. These styles also show tighter distributions with fewer low-similarity cases, suggesting that both humans and the language model draw on a more shared set of visual cues when describing these styles.

In contrast, styles such as elegant, dynamic, and sleek show lower median similarity scores and broader distributions with heavier lower tails, reflecting greater variation in participant interpretations and weaker alignment with generated captions (See Appendix \ref{appx:wordcloud} for examples of stated semantic associations across style words). Hartigan's dip tests indicated no evidence of multimodality in cosine similarity distributions across wheel designs for any style word (all $p > 0.67$), suggesting that semantic alignment varies continuously rather than forming discrete clusters. This pattern suggests that interpretations of certain styles are more subjective or context-dependent, making them harder to capture reliably through text-based descriptions alone. 

Overall, these results indicate that LLM-generated descriptions align more closely with human judgments for styles grounded in intuitive and widely recognized visual features. In contrast, alignment is weaker for more abstract styles, where interpretation depends on subjective or contextual factors that extend beyond the visual product domain.
\section{Discussion}
Our findings demonstrate that both designer-informed and computationally-derived visual features and consumer interpretations play complementary roles in understanding consumer perceptions of wheel aesthetics. We also reveal that the degree of differentiability in people's aesthetic perceptions varies depending on the specific style attribute. Below we discuss how this framework can be applied in product design workflows and elaborate on potential extensions of the computational framework to better capture and interpret consumer taste.

\subsection{Applying the Framework to Product Design}
%What does it take to adapt and apply this framework to other product domains?
The integrated framework presented here demonstrates clear potential for enabling product designers to understand, critique, and apply insights derived from human-centered analysis of consumer taste throughout the design process. Rather than treating consumer research as a late-stage validation activity, the framework is intended to support early and continuous engagement with perceptual signals, and allow designers to reflect on emerging concepts during ideation, exploration, and refinement. This approach aligns with core themes in AI and Design literature in HCI, which emphasize integrating design knowledge and AI through human-centered practices. We believe this goal can be achieved by combining designers' intuitive judgments with algorithmic generation and evaluation to support design processes as AI capabilities evolve.

By linking subjective perceptual judgments such as \textit{luxury} or \textit{sporty} to both human-derived design features and machine-extracted visual features, the framework externalizes latent consumer taste in an interpretable manner. Designers can systematically examine how formal variations influence aesthetic perceptions, identify gaps between intended and inferred meanings, and compare alternative design directions using shared perceptual constructs that are compatible with their design language. As designers generate sketches, parametric variations, or even early CAD models, the framework can surface how emerging designs align with or diverge from learned taste patterns over time. This enables design teams to maintain a continuous pulse on evolving consumer perceptions before designed artifacts reach higher levels of fidelity and become increasingly difficult to revise.

%By mapping perceptual judgments (e.g., luxury) to quantifiable human-derived features and machine-extracted features, the framework provides a flexible and adaptable structure for capturing and interpreting subjective preferences in diverse product domains that are heavily influenced by changing consumer tastes at different time scales. For instance, in fashion, features such as fabric texture, color schemes, and silhouette can be analyzed in relation to perceived high level semantic attributes such as cyberpunk or retrofuturism~\cite{davis2024fashioning}. The adaptability of our method extends beyond purely aesthetic-driven markets. In architectural and interior design, the framework could provide useful means of analyzing how spatial arrangements, lighting patterns, and material choices contribute to perceived ambiance and depth.

Although demonstrated in the context of car wheel design, the framework is structured to support adaptation to other product domains where visual form plays a significant role in shaping perception. In fashion, relationships between garment silhouettes, textures, and colors may be analyzed in relation to higher-level style constructs such as cyberpunk or retrofuturism~\cite{davis2024fashioning}. In architectural or interior design, spatial arrangements, lighting, and materials can similarly be examined in relation to perceived ambiance or depth. We believe the framework is most effective in domains where designers retain meaningful control over visual attributes that can be iteratively explored.

At the same time, the utility of the framework depends on the structure of the design space. One potential challenge in applying the framework lies in identifying appropriate domain-specific visual features and high-level style descriptors that meaningfully contribute to consumer perception. Unlike car wheels, which exist within a constrained design space with relatively isolated aesthetic considerations, other product categories (e.g., laptops or furniture with unibody construction) may have more interdependent and integrated design elements---rendering the framework less useful. In such cases, the framework may function more as a reflective lens for understanding how subtle design decisions shape perception.

Overall, the framework is not intended to optimize designs toward consumer preferences, but to support informed creative alignment and divergence. By rendering consumer taste as an interpretable and evolving signal, it enables designers to decide when to align with prevailing perceptions and when to deliberately depart from them, positioning computational analysis as a resource for creative judgment rather than a replacement for it.

\subsection{Extending the Framework}
Future research could explore domain-specific adaptations, refining feature representations and incorporating advanced machine learning techniques to enhance predictive accuracy. Additionally, expanding the framework to assess isolated product components (e.g., wheel) relative to its expected immediate context---including partial or full views of the product (e.g., body of the car), and even branding information---could further enhance our ability translate perceptual judgments into concrete product parameters in ways that are robust to these contextual influences.

Another extension could include examining the relationship between psycholinguistic properties of keywords such as imageability, concreteness, or context availability and people's exposure to product marketing to better understand their effect on our ability to interpret the results~\cite{altarriba1999concreteness}. In this case, highly imageable and concrete keywords tend to foster stronger visual associations, whereas less concrete language can obscure the intended stylistic impact and thereby reduce the precision with which our framework captures consumer perception. Our results suggest a need for more nuanced understanding of the complex interplay between human perception of aesthetics and linguistics. For instance, we suspect that consumers are predisposed to link wheel products to keywords like \textit{sporty} and \textit{classic} as they would often appear in marketing terminology for automotive products. On the contrary, the context for interpreting \textit{aerodynamic} and \textit{sleek} aesthetics, which exhibited relatively low discriminability in people's subjective judgments, may not have been specific to the automotive domain. For example, \textit{sleek} could describe a smooth, streamlined form, such as the surface of a high-speed train or a polished ceramic sculpture. The term \textit{aerodynamic} is considered broadly in different transportation products (e.g., train, fighter jet), but also evoke engineering knowledge that may be in conflict with certain perceived aesthetic qualities of a wheel (e.g., shape of a wind turbine blade vs. disc-shaped wheels).

Finally, our framework currently treats consumer taste as an aggregated perceptual signal and does not account for stable differences across individuals or groups, such as demographic background, experience, or attitudinal orientations. Future extensions could integrate people-centered information with linguistic analyses to move beyond population-level trends and better capture how shared perceptual constructs are interpreted differently across contexts and consumer segments~\cite{hakimi2023machine,chen2024understanding}. This extension would enable the framework to support a more nuanced understanding of aesthetic judgment and variation in perception, strengthening its value for creative decision-making in design.

\subsection{Limitations}
This work has several limitations that should be considered when interpreting the findings. First, our regression analyses rely on ordinary least squares (OLS) models applied to Bradley–Terry (BT) scores, which are derived estimates rather than directly observed continuous measures. While this approach is common in prior work, it assumes linear relationships and homoscedastic residuals, which may not fully capture the underlying structure of preference judgments or account for dependencies introduced during pairwise comparisons. Moreover, designer-informed visual features were aggregated as proportions of annotations per wheel, which smooths over potential disagreement among annotators and obscures feature co-occurrence patterns. Similarly, computationally derived low-level visual features capture measurable image properties but may omit higher-level perceptual cues that participants attend to when making aesthetic judgments. As a result, the feature sets used here should be understood as partial representations of the visual design space rather than exhaustive descriptions.

Second, our semantic alignment analysis depends on LLM-generated image captions and Sentence-BERT embeddings as proxies for visual interpretation. Although these methods enable scalable comparison between textual descriptions, they introduce modeling assumptions that may bias similarity estimates toward dominant linguistic patterns learned during pretraining. Cosine similarity reflects alignment in embedding space rather than perceptual similarity per se, and differences in participants’ verbal expressiveness or vocabulary may influence alignment scores independently of their visual perceptions.

Third, our interpretation of response variability and unimodality is limited by sample size and the resolution of the distributions. While Hartigan's dip test provides a principled way to assess unimodality, it does not distinguish between randomness and structured subgroup differences on its own. Consequently, conclusions about consensus versus heterogeneity should be interpreted as indicative rather than definitive, and future work could incorporate clustering or mixture modeling to more directly identify latent subgroups.

Finally, our analyses focus on a fixed set of wheel designs and nine predefined style keywords. These choices constrain the generalizability of the results to other product categories, design domains, or aesthetic vocabularies. Participants' interpretations of style terms are also shaped by cultural background and prior experience, which were not explicitly modeled in this study. Future work could examine how these factors interact with visual features and semantic interpretations over a broader range of stimuli and populations.
\section{Conclusion}
In this paper, we introduced a computational framework of consumers' aesthetic taste in product appearance, an essential component of creative expression. We applied our framework in the context of designing automotive products where capturing and modeling trends in consumer preferences is particularly challenging due to long development cycles. By demonstrating the viability of our approach in car wheels, we establish a foundation for its broader adoption across industries where consumer taste and aesthetic preference play a critical role. As digitalization of design processes continues to prevail, computational frameworks such as ours will be increasingly valuable in enabling designers and decision-makers to anticipate consumer responses. We hope our framework will give companies greater confidence in developing creative designs that resonate with evolving consumer aesthetic preferences.

\section{Acknowledgments}
Anonymized for review.

% \begin{acks}
% \end{acks}

\bibliographystyle{ACM-Reference-Format}
\bibliography{dis2026}

\appendix
\section{Appendix}
\label{appendix}
%\subsection{Style Keywords}
%\label{appx:stylekeywords}
%\begin{table}[h]
%    \centering
%    \begin{tabular}{cccc}
%        \textbf{Aerodynamic} & Aggressive & Bold & Chic\\
%        \textbf{Classic} & Complex & Contemporary & \textbf{Dynamic}\\
%        \textbf{Elegant} & Efficient & Emotional & \textbf{Futuristic}\\
%        Innovative & \textbf{Luxury} & Modern & Polished\\
%        Refined & \textbf{Rugged} & Sculpted & \textbf{Sleek}\\
%        \textbf{Sporty} & Streamlined & Strong & Stylish\\
%        Tough & & &
%    \end{tabular}
%    \caption{25 style keywords identified in prior study~\cite{jeon2024weaving}. Final selection of nine style keywords are bolded.}
%    \Description{Table listing 25 style keywords identified in a formative study of wheel aesthetics. The keywords are presented in grouped columns. Nine keywords are visually emphasized in bold %to indicate the final selection used in subsequent analyses, including classic, elegant, sporty, luxury, rugged, dynamic, futuristic, sleek, and aerodynamic.}
%    \label{tab:stylekeywords}
%\end{table}

\subsection{Text Prompt for Generating Wheel Captions with OpenAI GPT-5.2 Model}\label{appx:prompt}
\begin{lstlisting}[style=prompt]
Describe this car wheel image in detail. Write your description in 5-7 full sentences without using bullet points. Include information about the wheel design, style, finish, spoke pattern, and any other notable features
\end{lstlisting}

\subsection{Distribution of Bradley-Terry Scores}
To assess the normality of the Bradley-Terry scores, the Shapiro-Wilk test was conducted for each design attribute (style keyword). The results indicate that the distributions of the \textit{classic} ($p = 0.23050$), \textit{futuristic} ($p = 0.73634$), and \textit{rugged} ($p = 0.07764$) attributes do not significantly deviate from normality ($p > 0.05$). However, the remaining attributes, including \textit{aero} ($p = 0.00045$), \textit{dynamic} ($p = 0.00022$), \textit{elegant} ($p < 0.00000$), \textit{luxury} ($p < 0.00000$), \textit{sleek} ($p = 0.00002$), and \textit{sporty} ($p = 0.00013$), exhibit significantly low $p$-values ($p < 0.05$), suggesting departures from normality.

\begin{table*}[h] 
    \centering
    \begin{tabular}{lcccccccc}
        \toprule
        \textbf{} & \textbf{mean} & \textbf{std} & \textbf{min} & \textbf{25\%} & \textbf{50\%} & \textbf{75\%} & \textbf{max} & \textbf{range} \\
        \midrule
        aero        & 0.00036  & 0.388392 & -1.261099 & -0.141786 & 0.076713  & 0.258842  & 0.897681  & 2.15878  \\
        classic     & 0.000847 & 0.656420 & -2.083226 & -0.416937 & 0.042958  & 0.404936  & 1.310534  & 3.39376  \\
        dynamic     & 0.000883 & 0.653757 & -2.328774 & -0.346055 & 0.121833  & 0.436877  & 1.164117  & 3.492891 \\
        elegant     & 0.000592 & 0.526522 & -2.346936 & -0.162169 & 0.115825  & 0.317189  & 0.799268  & 3.146204 \\
        futuristic  & 0.000772 & 0.578325 & -1.681770 & -0.378392 & 0.052493  & 0.325395  & 1.508763  & 3.190533 \\
        luxury      & 0.000804 & 0.584391 & -2.624126 & -0.249351 & 0.092800  & 0.389529  & 0.800820  & 3.424946 \\
        rugged      & 0.000478 & 0.451417 & -0.912717 & -0.344027 & -0.077797 & 0.298714  & 1.059845  & 1.972562 \\
        sleek       & 0.000490 & 0.452063 & -1.902704 & -0.141780 & 0.064055  & 0.300412  & 0.700480  & 2.603184 \\
        sporty      & 0.001074 & 0.697550 & -2.563169 & -0.380285 & 0.134286  & 0.430761  & 1.063796  & 3.626965 \\
        \bottomrule
    \end{tabular}
    \caption{Descriptive statistics for Bradley-Terry scores across different design style keywords.}
    \Description{Table reporting descriptive statistics for Bradley–Terry scores across nine design style keywords. For each style, the table lists mean, standard deviation, minimum, first quartile, median, third quartile, maximum, and range values. The statistics summarize the distribution and variability of preference scores across styles.}
    \label{tab:BTscores}
\end{table*}
\newpage
\subsection{Participant Stated Semantic Associations with Style Keywords in Response to Wheel Stimuli}\label{appx:wordcloud}
\begin{figure}[h]  % 't' for top placement
\centering
\includegraphics[width=\linewidth]{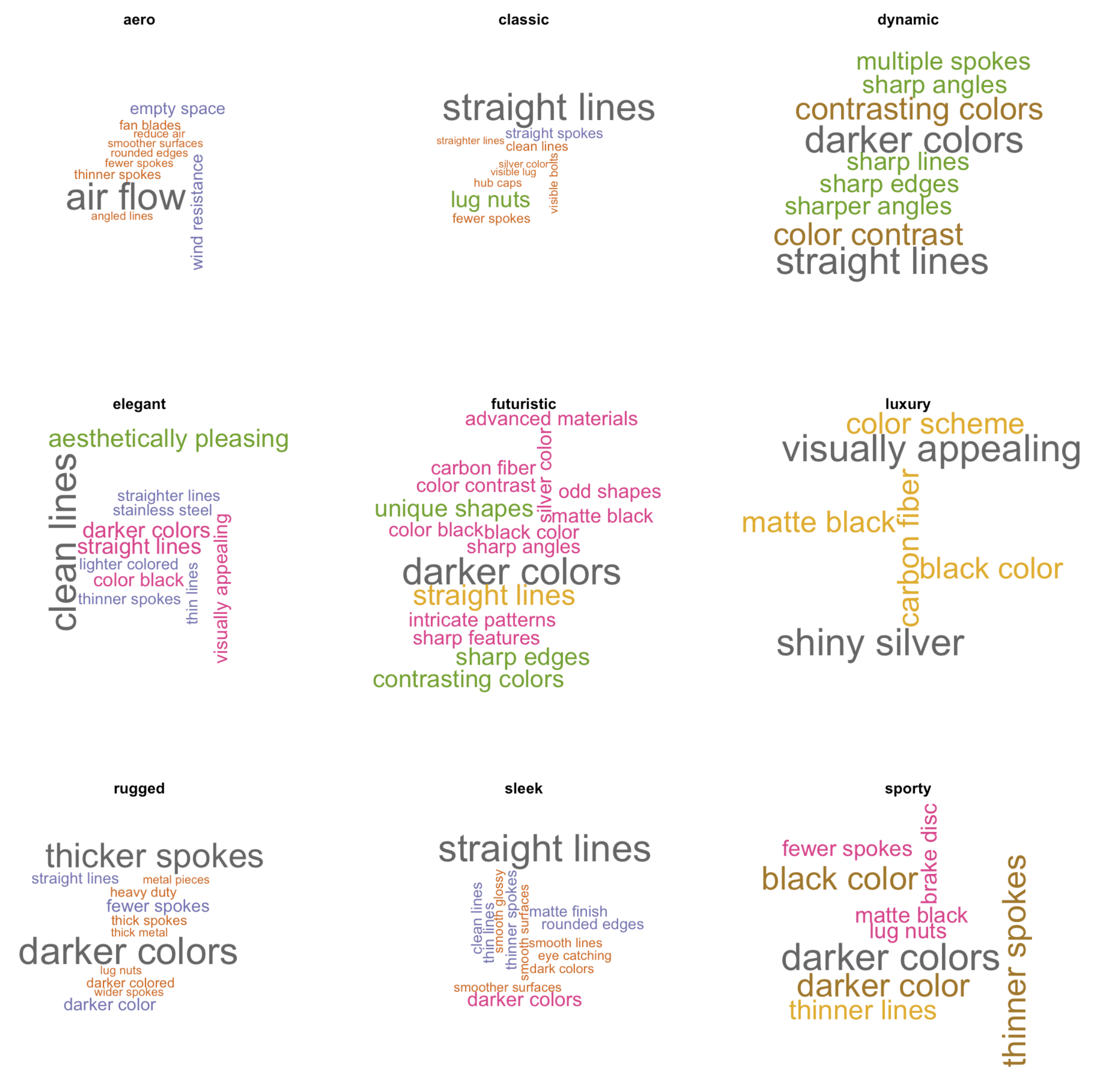} 
\caption{Wordcloud visualizations generated from commonly occurring bi-grams from participant descriptions of wheel visual features. Semantic associations elicited by different aesthetic styles include references to domain-relevant design language such as spoke shape, whereas \textit{aero} style includes include semantic associations external to the wheel design domain such as "fan blades" and "air flow".}
\Description{Set of word cloud visualizations showing commonly occurring bi-grams from participant descriptions of wheel visual features, grouped by nine aesthetic styles. Larger words indicate higher frequency. Styles such as classic, elegant, and luxury emphasize terms related to straight lines, clean forms, and material finishes, while aero includes terms related to airflow and fan blades, and rugged emphasizes thicker spokes and darker colors.}
\label{fig:wordcloud}
\end{figure}

%\begin{figure*}[h]
%    \centering
%    \includegraphics[width=0.8\textwidth]{fig/results_BTdistribution.png}
%    \caption{Frequency distribution of BT scores.}
%    \label{fig:results_BTdistribution}
%\end{figure*}

%\begin{figure}[t]
%    \centering
%    \includegraphics[width=0.9\linewidth]{fig/appx_pca.png}
%    \caption{Visualization of K-means clusters after dimensionality reductions with Principle Components Analysis.}
%    \label{fig:appx_pca}
%\end{figure}

\newpage
\subsection{Influence of Split Type on Aesthetic Perceptions}

\begin{figure}[h]
    \centering
    \includegraphics[width=1\linewidth]{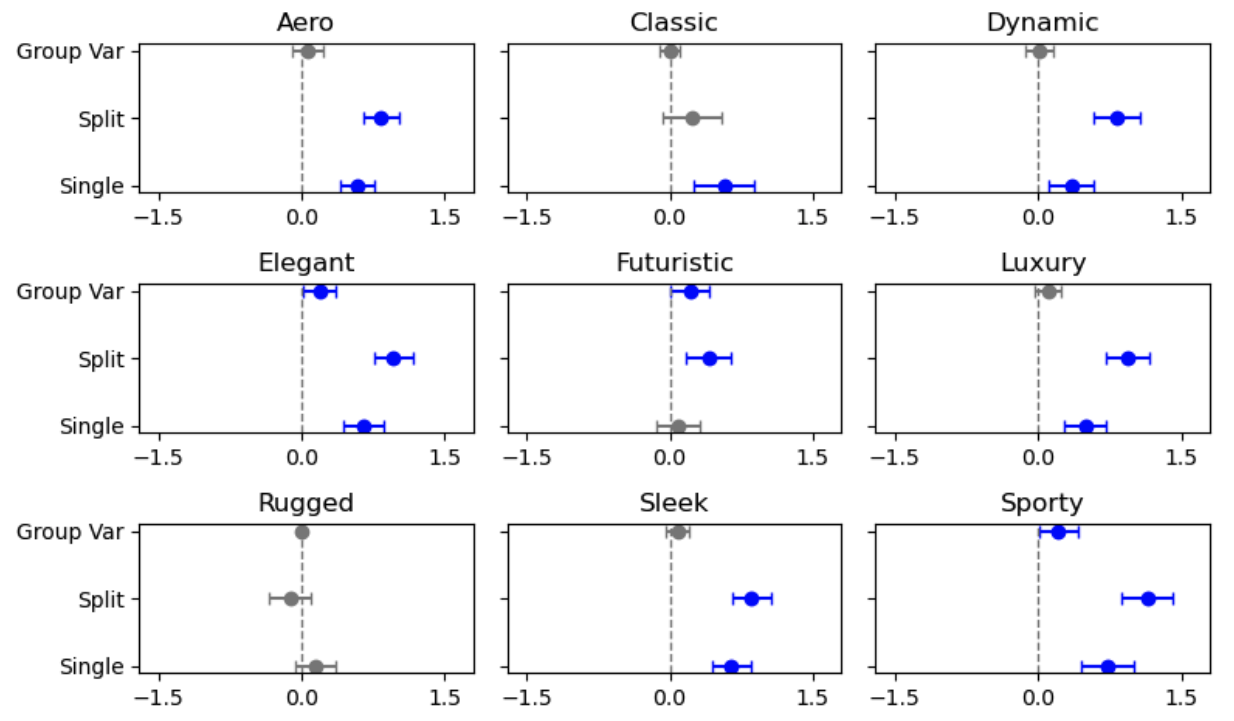}
    \caption{Influence of split type on aesthetic perceptions.}
    \Description{Grid of regression plots showing the influence of split type on aesthetic perceptions across nine style keywords. Each subplot displays estimated coefficients for different split configurations, including single and split spokes, with points and horizontal error bars representing effect sizes and confidence intervals. A vertical dashed line indicates zero effect, highlighting which split types are positively, negatively, or not significantly associated with each style.}
    \label{fig:appx_splittype}
\end{figure}

\subsection{Designer-Informed Visual Features}\label{appx:wheelfeatures}
\begin{table*}[t]
    \centering
    \small
    \renewcommand{\arraystretch}{1.2} % Adjust for extra vertical spacing
    \setlength{\tabcolsep}{5pt}       % Adjust column separation
    % Columns 1-4 centered horizontally, column 5 left aligned, all vertically centered
    \begin{tabular}{
        >{\centering\arraybackslash}m{2cm}
        >{\centering\arraybackslash}m{2cm}
        >{\centering\arraybackslash}m{2cm}
        >{\centering\arraybackslash}m{2cm}
        >{\arraybackslash}m{9cm}
    }
        \textbf{Type} & \textbf{Category} & \textbf{Features} & \textbf{Example} & \textbf{Description} \\
        \toprule
        Directional & --- & --- 
        & \includegraphics[width=0.9\linewidth]{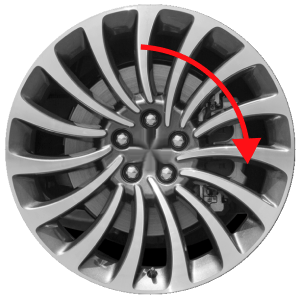}
        & Curvature or angle of spokes trend (counter) clock-wise direction \\
        \midrule
        Split Type & Zero & ---
        & \includegraphics[width=0.9\linewidth]{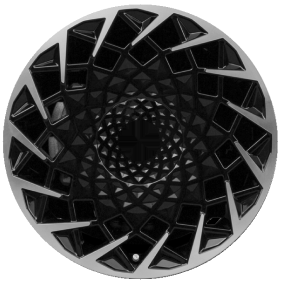}
        & No spokes. Typically a disc shape \\
        & Single & ---
        & \includegraphics[width=0.9\linewidth]{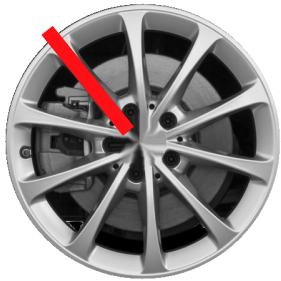}
        & Simple straight spokes \\
        & Split & Double
        & \includegraphics[width=0.9\linewidth]{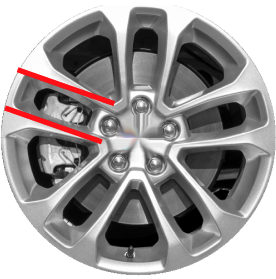}
        & Pair of parallel spokes \\
        & & Triple
        & \includegraphics[width=0.9\linewidth]{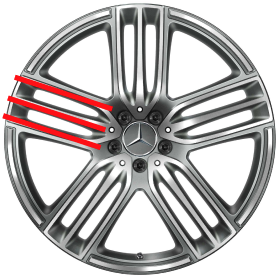}
        & Triplets of parallel spokes \\
        & & V
        & \includegraphics[width=0.9\linewidth]{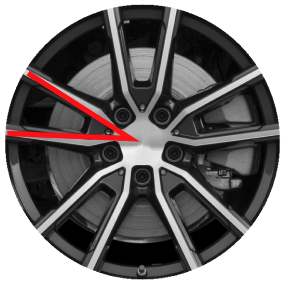}
        & Pair of spokes opening up in a V shape. Spokes are joined at the center. \\
        & & Y
        & \includegraphics[width=0.9\linewidth]{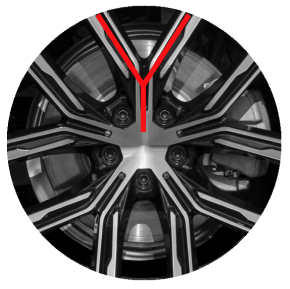}
        & Pair of spokes branching out in a Y shape from the stem of a spoke \\
        & & Complex
        & \includegraphics[width=0.9\linewidth]{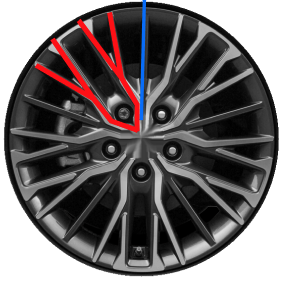}
        & Complex multiple branching pattern (e.g., Y-spoke branching out of a V-spoke) \\
        \midrule
        Arrangement & & Offset
        & \includegraphics[width=0.9\linewidth]{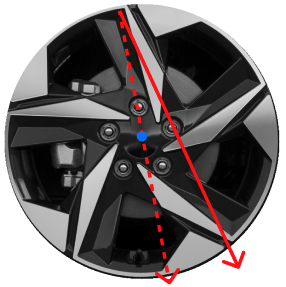}
        & Spokes that are positioned at an angle or offset from the center \\
        & & Double-stacked
        & \includegraphics[width=0.9\linewidth]{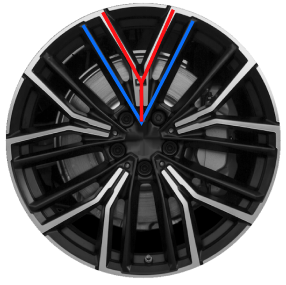}
        & Different spoke designs layered on top of each other \\
        & & Hollowed
        & \includegraphics[width=0.9\linewidth]{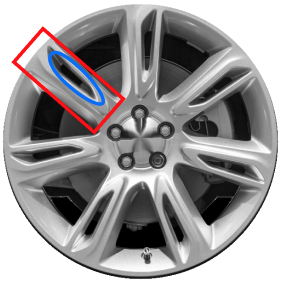}
        & Spokes that appear to be split but joined near the end of the rim \\
        & & Indented
        & \includegraphics[width=0.9\linewidth]{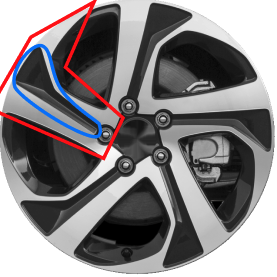}
        & Spokes that have narrow recessed portions along the length of the spoke \\
        \bottomrule
    \end{tabular}
    \caption{Overview of designer-informed visual features and corresponding images organized by type, category, and individual features.}
    \Description{Table illustrating designer-informed visual feature categories for wheel design, organized by type and subcategory. The table includes example wheel images annotated to show presence of directionality, split types such as zero, single, double, triple, V, Y, and complex spoke patterns, as well as arrangement features including offset, double-stacked, hollowed, and indented spokes. Short textual descriptions explain each feature type.}
    \label{tab:designerfeatures}
\end{table*}

\end{document}